\documentclass[a4paper,11pt]{article}
\usepackage{jheppub} 
\usepackage{xcolor}
\usepackage{physics}
\usepackage{xspace}
\usepackage{mathrsfs}
\usepackage{soul}
\usepackage{mathtools}
\usepackage{tensor}
\usepackage{tabularx}
\usepackage{cancel}
\usepackage{booktabs}

\makeatletter
\g@addto@macro\bfseries{\boldmath}
\makeatother

\newcommand{\A}{\mathcal{A}}
\newcommand{\B}{\mathcal{B}}
\newcommand{\C}{\mathcal{C}}
\newcommand{\D}{\mathcal{D}}

\preprint{Imperial-TP-2024-CH-01, UUITP-04/24}

\title{Charges and topology in linearised gravity}

\author[a]{Chris Hull,}
\author[a]{Maxwell L. Hutt,}
\author[a,b]{Ulf Lindstr\"{o}m}

\affiliation[a]{The Blackett Laboratory, Imperial College London, Prince Consort Road, London, SW7 2AZ, UK}
\affiliation[b]{Theoretical Physics and Centre for Geometry and Physics, Uppsala University, SE-751 20, Uppsala, Sweden}

\emailAdd{c.hull@imperial.ac.uk, m.hutt22@imperial.ac.uk, ulf.lindstrom@physics.uu.se}

\abstract{Covariant conserved 2-form currents for linearised gravity are constructed by contracting the linearised curvature with conformal Killing-Yano tensors. The corresponding conserved charges were originally introduced by Penrose and have recently been interpreted as the generators of generalised symmetries of the graviton. We introduce an off-shell refinement of these charges and find the relation between these improved Penrose charges and the linearised version of the ADM momentum and angular momentum. If the graviton field is globally well-defined on a background Minkowski space then some of the Penrose charges give the momentum and angular momentum while the remainder vanish.
We consider the generalisation in which the graviton has Dirac string singularities or is defined locally in patches, in which case the conventional ADM expressions are not invariant under the graviton gauge symmetry in general. We modify them to render them gauge-invariant and show that the Penrose charges give these modified charges plus certain magnetic gravitational charges. We discuss properties of the Penrose charges, generalise to toroidal Kaluza-Klein compactifications and check our results in a number of examples.
}

\begin{document}
\maketitle
\flushbottom

\section{Introduction}

In a remarkable paper \cite{Penrose1982Quasi-localRelativity} Penrose introduced a conserved 2-form current $Y$ in linearised gravity of the form
\begin{equation}
    Y[K]_{\mu\nu} = R_{\mu\nu \alpha\beta} K^{\alpha\beta}
\end{equation}
where $R_{\mu\nu\alpha\beta}$ is the linearised curvature tensor for metric fluctuations about a background Minkowski spacetime. This is conserved, i.e.
\begin{equation}\label{eq:Y_conserved_intro}
    \partial^{\mu} Y[K]_{\mu\nu} = 0,
\end{equation}
if the vacuum Einstein equations hold, i.e. $R_{\mu\nu}=0$, and $K_{\alpha\beta}$ is a 2-form satisfying
\begin{equation}\label{eq:CKY_eqn_intro}
    \partial_\lambda K_{\mu\nu} = \tilde{K}_{\lambda\mu\nu} + 2 \eta_{\lambda[\mu}\hat{K}_{\nu]}
\end{equation}
where
\begin{equation}\label{eq:Ktilde_Khat_def_intro}
    \tilde{K}_{\lambda\mu\nu} = \partial_{[\lambda}K_{\mu\nu]} \qq{and} \hat{K}_\mu = \frac{1}{d-1} \partial^\nu K_{\nu\mu}
\end{equation}
Tensors satisfying (the covariant version of) this equation are known as \emph{conformal Killing-Yano} (CKY) tensors \cite{Tachibana1969OnSpace,Kashiwada1968}, so that the tensors $K$ satisfying eq.~\eqref{eq:CKY_eqn_intro} are the CKY tensors for Minkowski space.
Penrose gave a twistorial interpretation of this equation in four dimensions \cite{Penrose1982Quasi-localRelativity}.

While Penrose focused on four dimensions, his construction extends to $d$ dimensions.
From eq.~\eqref{eq:Y_conserved_intro}, $\star Y$ is a closed $(d-2)$-form and the integral of this over a $(d-2)$-surface $\Sigma$ defines a conserved charge 
\begin{equation}\label{eq:Q[K]_intro}
    Q[K] = \int_{\Sigma} \star Y[K]
\end{equation}
This charge is unchanged under deformations of $\Sigma$ that do not cross any points at which $R_{\mu\nu}\neq0$ and the charge provides a measure of the amount of mass/energy contained within $\Sigma$.
Penrose interpreted these charges as giving covariant expressions for the total momentum and angular momentum contained within $\Sigma$. 
He argued that for each Minkowski space Killing vector there is a corresponding CKY tensor and that the charges $Q[K]$ for these CKY tensors give covariant expressions for the total momentum (corresponding to the translation Killing vectors) and angular momentum (corresponding to the Lorentz Killing vectors).

Penrose went on to generalise his construction to curved spacetime. General spacetimes do not admit Killing vectors or CKY tensors, but, to construct $Q[K]$ in eq.~\eqref{eq:Q[K]_intro}, $\star Y$ is only needed on the surface $\Sigma$, not over the whole spacetime. Penrose constructed a 2-form $\star Y$ on $\Sigma$ using `surface twistors' and proposed that this gives a quasi-local definition of momentum and angular momentum in general relativity. On taking $\Sigma$ to be at null infinity he obtained the BMS momentum together with an angular momentum.

Penrose's charges have been rediscovered a number of times (see, e.g., Refs.~\cite{Jezierski:2002mn, Jezierski:2014gka}), and have been considered recently in Refs.~\cite{Hinterbichler2023GravitySymmetries, Benedetti2022GeneralizedGraviton, Benedetti2023GeneralizedGravitons, BenedettiNoether, Gomez-Fayren2023CovariantRelativity} in the context of higher-form symmetries of the graviton. 
The Penrose charges can be viewed as gauge-invariant topological operators defined on codimension-2 cycles that generate 1-form symmetries \cite{Gaiotto2015GeneralizedSymmetries}.

Not all the CKY tensors correspond to Killing vectors. Namely, a particular class of CKY tensors are the \emph{Killing-Yano} (KY) tensors which satisfy eq.~\eqref{eq:Ktilde_Khat_def_intro} with $\hat{K}_\mu=0$, and do not have corresponding Killing vectors. 
For example, in four dimensions the space of Killing vectors is 10-dimensional while that of CKY tensors is 20-dimensional.
This raises the question of the significance of the Penrose currents corresponding to the KY tensors. 
Penrose avoided this mismatch by imposing a hermiticity condition on the (twistor form of) his charges that left 10 real charges. This eliminated certain gravitational analogues of magnetic charge and one of the aims of this paper is to revisit this correspondence if gravitational magnetic charges of the kind analysed in Ref.~\cite{HullYetAppear} are included.
In Ref.~\cite{Hinterbichler2023GravitySymmetries}, all the Penrose charges for linearised gravity in four dimensions were associated with certain parameters (e.g. a NUT parameter) in a linearised solution, providing some insight into their significance in that case.

In this paper we consider the Penrose charges for the $d$-dimensional free graviton theory in Minkowski space with the Fierz-Pauli action.
We construct an off-shell refinement of the Penrose currents with extra terms involving the linearised Ricci tensor that vanish on-shell. 
For the case in which $K$ is a KY tensor, the improved Penrose currents are identically conserved (without using field equations) and are in fact the currents constructed by Kastor and Traschen \cite{Kastor2004ConservedTensors}. 
We then derive a precise relation between the charges that are given by integrating the improved Penrose currents and the momentum and angular momentum that arise from the linearisation of the ADM construction \cite{ADM,Abbott1982StabilityConstant}. 
If the graviton field is non-singular and defined on the whole of Minkowski space, then the relation is straightforward and the Penrose charges for one class of CKY tensors give the ADM charges while the remainder vanish.
However, if the graviton field is not globally defined in the sense that it has Dirac string singularities or is defined in patches with transition functions involving gauge transformations, then topological or magnetic charges for gravity of the kind recently constructed in Ref.~\cite{HullYetAppear} can arise. In particular, if the graviton field is not globally defined, then total derivative contributions become important. 
We show that the standard expression for each ADM charge is only gauge-invariant up to a surface term, and this can be non-zero if the graviton is not globally defined.
As a result, the standard ADM expressions are in general not gauge-invariant if the graviton field is not globally defined.
We find a surface-term modification of the standard ADM expressions that is fully gauge invariant under these circumstances.
We then relate the Penrose charges, which are manifestly gauge-invariant, to these covariant improved ADM charges together with certain gravitational magnetic charges.

The structure of the paper is as follows. In section \ref{sec:ADMReview}, we review the theory of linearised gravity and outline the construction of the ADM charges and the  gravitational magnetic charges introduced in Ref.~\cite{HullYetAppear}. In section \ref{sec:Penrose2Form}, we discuss Penrose's 2-form current in more detail and derive an off-shell refinement of it which is used in section~\ref{sec:PenADM}, where the relationship between the Penrose and ADM charges is derived. We then analyse the charges constructed by integrating the Penrose 2-form current and the dual charges constructed by integrating its Hodge dual. 
We do this for spacetime dimensions $d>4$ in sections \ref{sec:PenroseElectricRelation} and \ref{sec:d-3_form_charges}. The situation in $d=4$ dimensions is different  and is analysed in section \ref{sec:four_dimensions}. 
In section \ref{sec:KaluzaKlein}, we discuss the charges that arise when the background Minkowski space is replaced by a product of Minkowski space with a torus, allowing the Kaluza-Klein reduction of the linearised theory. These form a  more general set of charges than those that arise in Minkowski space. We first derive the charges in the dimensionally reduced theory, then give their uplift to the higher dimensional theory. Section \ref{sec:examples} gives examples of solutions of the linearised theory, calculates their charges and checks the relationship between the Penrose charges and the ADM and magnetic charges. Finally, in section \ref{sec:conclusion}, we summarise our results and discuss their implications.

\section{Linearised gravity and its conserved charges}
\label{sec:ADMReview}

We study the spin-2 free graviton field $h_{\mu\nu}$ in $d$-dimensional Minkowski space with global Cartesian coordinates $x^\mu$ and  with Minkowski metric $\eta_{\mu\nu}=\text{diag}(-1,1,1\dots,1)$.
We will later consider configurations in which the graviton field is singular at the locations of certain sources, and in that case we will restrict to the space $\mathcal{M}$ given by Minkowski space with  points or regions removed, so that $h_{\mu\nu}$ is a non-singular field on $\mathcal{M}$.

The graviton is a symmetric tensor with gauge transformation 
\begin{equation}\label{eq:h_gauge_transformation}
    h_{\mu\nu} \to h_{\mu\nu} + \partial_\mu \zeta_\nu + \partial_\nu \zeta_\mu
\end{equation}
The invariant field strength is the linearised Riemann tensor
\begin{equation} \label{eq:DefRiemann}
    R_{\mu\nu\alpha\beta} = \partial_\alpha \Gamma_{\mu\nu|\beta} - \partial_\beta \Gamma_{\mu\nu|\alpha}
\end{equation}
where\footnote{We will used square brackets to indicate anti-symmetrisation with unit weight, e.g. $T_{[\mu\nu]} = \frac{1}{2}(T_{\mu\nu} - T_{\nu\mu})$.}
\begin{equation} \label{eq:DefConnection}
    \Gamma_{\alpha\beta|\mu} = \partial_{[\alpha}h_{\beta]\mu}
\end{equation}
The Fierz-Pauli field equation with source is 
\begin{equation}\label{eq:field_equation}
    G_{\mu\nu}= T_{\mu\nu}
\end{equation}
where $G_{\mu\nu}$ is the linearised Einstein tensor
\begin{equation}
    G_{\mu\nu} = R_{\mu\nu} - \frac{1}{2} \eta_{\mu\nu} R
\end{equation}
with $R_{\mu\nu}$ and $R$  the linearised Ricci tensor and scalar respectively.
The source is a symmetric tensor that is conserved, 
\begin{equation}\label{eq:T_conserved}
    \partial ^\mu T_{\mu\nu}=0\, .
\end{equation}
In general, solutions of this free theory  need not arise as the linearisation of solutions of the full non-linear Einstein theory.

The Minkowski space Killing vectors $k^\mu$ satisfy
\begin{equation}\label{killlin}
  \partial_{(\mu}k_{\nu)}=0
\end{equation}
and are given by
\begin{equation}\label{eq:Killing_vectors}
    k_\mu = V_\mu + \Lambda_{\mu\nu} x^\nu
\end{equation}
where $V$ is a constant 1-form and $\Lambda$ is a constant 2-form, corresponding to translations and Lorentz transformations respectively.
Then for any Killing vector $k$,
\begin{equation}\label{eq:j[k]}
    j[k]_\mu = T_{\mu\nu}k^\nu
\end{equation}
is a conserved current
\begin{equation}  
    \partial^\mu j _\mu=0
\end{equation}
using eqs.~\eqref{eq:T_conserved} and \eqref{killlin}.

Using the field equation \eqref{eq:field_equation}, the current can be written in terms of the graviton field as
\begin{equation}\label{eq:jh[k]}
    j[k]_\mu = G_{\mu\nu}k^\nu
\end{equation}
which is a total derivative \cite{Abbott1982StabilityConstant},
\begin{equation}
    j[k]_\mu =\partial ^\nu J_{\mu\nu}
\end{equation}
where
\begin{equation} \label{eq:J[k]_def}
    J[k]^{\mu\nu} = \partial_\sigma \mathcal{K}^{\mu\nu|\rho\sigma} k_\rho - \mathcal{K}^{\mu\sigma|\rho\nu} \partial_\sigma k_\rho
\end{equation}
with 
\begin{equation} \label{eq:DefK}
    \mathcal{K}^{\mu\nu|\rho\sigma} = -3\eta^{\mu\nu\alpha|\rho\sigma\beta} h_{\alpha\beta}
\end{equation}
and
\begin{equation}
    \eta^{\mu\nu\rho|\alpha\beta\gamma} = \eta^{\alpha\sigma} \eta^{\beta\kappa} \eta^{\gamma\lambda} \delta^\mu_{[\sigma} \delta^\nu_\kappa \delta^\rho_{\lambda]}
\end{equation}

For a $(d-1)$-dimensional hypersurface $S$ with boundary $\Sigma_{d-2}$ we define the charge
\begin{equation}\label{qkis}
    Q[k] = \int_S j[k]_\mu \dd{\Sigma^\mu}
\end{equation}
which can be rewritten as a surface integral over the boundary
\begin{equation}\label{qkisb}
    Q[k] = \frac{1}{2} \int_{\Sigma_{d-2}} J[k]_{\mu\nu} \dd{\Sigma^{\mu \nu}}
\end{equation}
If $S$ is taken to be a hypersurface of fixed time with $\Sigma_{d-2}$ the $(d-2)$-sphere at spatial infinity, then $Q[k]$ for a Killing vector of the form \eqref{eq:Killing_vectors} is a conserved charge giving the linearised ADM momenta $P_\mu$ and angular momenta $L_{\mu\nu}$,
\begin{equation}\label{eq:ADM_charge_definition}
    Q[k] = V_\mu P^\mu + \frac{1}{2} \Lambda_{\mu\nu} L^{\mu\nu}
\end{equation}
If $S$ is taken to be a region of a hypersurface of fixed time with boundary $\Sigma_{d-2}$ then $Q[k]$ is conserved if $\Sigma_{d-2}$ is contained within a region without sources (where $j[k]$ vanishes).
Then $Q[k]$ gives the momentum and angular momentum in the $(d-1)$-dimensional region $S$ bounded by $\Sigma_{d-2}$. 
Note that this is different from Einstein gravity where the charge is only conserved if $\Sigma_{d-2}$ is taken to be the sphere at infinity and the fields satisfy suitable boundary conditions. 

The Killing vectors $k$ give \emph{invariances} of the theory: any field configuration is invariant under the gauge transformation \eqref{eq:h_gauge_transformation} in which $\zeta_\mu$ is a Killing vector. Then $Q[k]$ is the conserved charge corresponding to the invariance under eq.~\eqref{eq:h_gauge_transformation} with $\zeta_\mu = k_\mu$, and we will refer to it as an \emph{electric}-type charge for the graviton.
We will refer to $j[k]$ as the \emph{primary} current associated with the invariance and a 2-form current $J_{\mu\nu}$ with
\begin{equation}
    j_\mu = \partial^\nu J_{\mu \nu}
\end{equation}
will be referred to as a \emph{secondary} current.
Note that if $J$ is such a secondary  current, then so is $J' = J + \dd^\dagger L$ for any 3-form $L$.

In Ref.~\cite{HullYetAppear}, \emph{magnetic} charges for linearised gravity were discussed.
These are all of the form 
\begin{equation}
    Q =  \frac{1}{2}  \int_{\Sigma_{d-2}} J _{\mu\nu}^{\text{mag}}\dd{\Sigma^{\mu \nu}}
\end{equation}
where the current is a total derivative
\begin{equation}\label{eq:div_of_3_form}
    J_{\mu\nu}^{\text{mag}}= \partial^\rho Z_{\mu\nu\rho}
\end{equation}
for some totally antisymmetric $Z_{\mu\nu\rho}$, so that $J^{\text{mag}}$ is automatically a conserved 2-form current. If $Z_{\mu\nu\rho}$ is a globally defined 3-form, the current is trivial and the charge is zero, so non-trivial magnetic charges only arise when this is not the case.

In the cases we consider here, $Z_{\mu\nu\rho}$ has a local expression in terms of the graviton field $h_{\mu\nu}$ and globally defined Killing vectors or Killing tensors.
If the components of $h_{\mu\nu}$ are non-singular functions defined over the whole of Minkowski space, so that $h_{\mu\nu}$ is a globally defined tensor on Minkowski space, then $\star J^{\text{mag}}$ is exact and the charge $Q=\int_\Sigma \star J^{\text{mag}}$ is zero.
Then, to obtain a non-trivial magnetic charge, it is necessary that 
$h_{\mu\nu}$ is not a globally defined tensor field on the entire Minkowski space.
Typically for the solutions to eq.~\eqref{eq:field_equation} with magnetic charges, the graviton field is not defined on the whole Minkowski space but instead on a space $\mathcal{M}$ which is Minkowski space with some points or regions removed, so that it can have non-trivial topology. (The regions removed from Minkowski space can be associated with the locations of magnetic sources \cite{HullYetAppear}.) 
Then to obtain non-trivial charges, $h_{\mu\nu}$ should be a field on $\mathcal{M}$ with a Dirac string singularity or it should be defined locally in patches of $\mathcal{M}$ with transition functions involving non-trivial gauge transformations of the form \eqref{eq:h_gauge_transformation}, giving a topologically non-trivial field configuration.
Although $h_{\mu\nu}$ need not be globally defined, the field strength $R_{\mu\nu\rho\sigma}$ is globally defined and gauge-invariant.
In  cases with magnetic charges, if we try to analytically extend such a  $h_{\mu\nu}$ defined in one patch to the whole of Minkowski space, we find Dirac string singularities.
As these charges are integrals of a total derivative, they are topological charges. See Ref.~\cite{HullYetAppear} for further discussion. 

In principle, any local 3-form $Z$ could be used to construct such a charge. Ref.~\cite{HullYetAppear} focused on the charges that arise as electric charges for the dual graviton theory introduced in Ref.~\cite{Hull2000}.
These charges arise from invariances of the dual graviton theory that correspond to Killing vectors $k$ in four dimensions or to generalised Killing tensors denoted $\kappa$ and $\lambda$ in $d>4$. 
In regions without magnetic sources for the dual graviton, some of these electric charges for the dual graviton can be dualised to the graviton theory where they become magnetic charges given by the integral of a total derivative. We now discuss these in more detail.

In four dimensions, these electric charges for the dual graviton result in a topological charge $\tilde Q[k]$ for the graviton theory for each Killing vector $k$, with currents $\tilde J[k]$ given by eq.~\eqref{eq:div_of_3_form} with
\begin{equation}\label{eq:Z[k]}
    Z[k]_{\mu\nu\rho} = - \epsilon_{\mu\nu\rho\sigma} k^\tau h\indices{_\tau^\sigma}
\end{equation}
giving dual momentum and angular momentum
\begin{equation}\label{eq:4d_dual_ADM_charges}
    \tilde{Q}[k] = V_\mu \tilde{P}^\mu + \frac{1}{2} \Lambda_{\mu\nu} \tilde{L}^{\mu\nu}
\end{equation}
Here $\tilde{P}^\mu$ is the linearised version of the dual momentum or NUT 4-momentum introduced for general relativity in Refs.~\cite{RamaswamyDualMass,AshtekarNUTMomenta} and $\tilde{L}_{\mu\nu}$ can be viewed as a  dual angular momentum charge.

In $d>4$, there are two types of magnetic charges for the graviton theory which correspond to electric charges for the dual graviton.
The first involves a rank-$(d-3)$ KY tensor $\lambda_{\mu_1\ldots \mu_{d-3}}$ which by definition satisfies
\begin{equation}\label{lkill}
    \partial_\rho \lambda_{\mu_1\mu_2 \dots \mu_{d-3}} - \partial_{[\rho} \lambda_{\mu_1\mu_2 \dots\mu_{d-3}]}=0
\end{equation}
Then a current $J[\lambda]_{\mu\nu}$  is defined by  eq.~\eqref{eq:div_of_3_form} with \cite{HullYetAppear}
\begin{equation}\label{eq:Z[lambda]}
    Z[\lambda]_{\mu\nu\rho} = (-1)^{d+1} \frac{4}{d-1} \tilde \lambda_{[\sigma \mu\nu }h_{\rho]}{}^ \sigma   
\end{equation}
where
\begin{equation} 
    \tilde{\lambda}^{\alpha\beta\gamma} = \frac{1}{(d-3)!} \epsilon^{\alpha\beta\gamma\mu_{1}\mu_{2}\ldots\mu_{d-3}} \lambda_{\mu_1\ldots \mu_{d-3}} 
\end{equation}
is a closed CKY tensor.\footnote{This is because the dual of a KY tensor is a closed CKY tensor (see Appendix~\ref{app:AppendixCKY}).}
For constant KY tensors $\lambda$, the charge 
\begin{equation}\label{qkisl}
    Q[\lambda] = \frac{1}{2} \int_{\Sigma_{d-2}} J[\lambda]_{\mu\nu} \dd{\Sigma^{\mu\nu}}
\end{equation}
arises as an electric charge for the dual graviton theory \cite{HullYetAppear}. 
The magnetic charge $\hat{P}_{\mu_1\dots\mu_n}$ is defined as the charge constructed from $J[\lambda]$ for constant $\lambda$, that is
\begin{equation}\label{eq:dual_ADM_charge_definition}
    Q[\lambda] = \int_{\Sigma_{d-2}} \star J[\lambda] = \frac{1}{n!} \lambda_{\mu_1\dots\mu_n} \hat{P}^{\mu_1\dots\mu_n}
\end{equation}
However, the integral of \eqref{eq:Z[lambda]} gives a conserved charge for any $(d-3)$-form $\lambda$, and in later sections this charge will arise for non-constant tensors $\lambda$. For non-constant $\lambda$, however, they cannot be straightforwardly interpreted as electric charges for the dual graviton due to the non-local relation of the graviton to its dual. Further discussion of this will appear in a forthcoming paper.

The other charge that arises as an electric charge for the dual graviton theory involves a generalised Killing tensor $\kappa_{\mu_1\dots \mu_{d-4} | \rho}$ which is in the $\text{GL}(d)$ representation corresponding to a Young tableau with one column of length $d-4$ and one of length $1$.
It satisfies the generalised Killing condition \cite{HullYetAppear}
\begin{equation}\label{kkill}
    \partial_{[\lambda} \kappa_{\mu_1 \dots \mu_{d-4}] | \rho} =0
\end{equation}
In this case, the secondary current is of the form \eqref{eq:div_of_3_form} with the 3-form $Z_{\mu\nu\rho}$ given by 
\begin{equation}\label{eq:Z[kappa]}
    Z[\kappa]_{\mu\nu\rho} = (-1)^{d} \frac{1}{d-1} \, \tilde{\kappa}_{\mu\nu\rho\sigma | \tau} h^{\sigma\tau}
\end{equation}
where
\begin{equation}\label{dukapis}
    \tilde{\kappa}^{\alpha\beta\gamma\delta | \nu} = \frac{1}{(d-4)!} \kappa_{\mu_1 \mu_2\ldots \mu_{d-4}|}{}^{\nu} \epsilon ^{\mu_{1}\mu_{2}\ldots\mu_{d-4} \alpha \beta \gamma \delta} 
\end{equation}
giving a current $J_{\mu\nu}[\kappa]$ and a charge $Q[\kappa]$.
These  do not play a role here as they do not correspond to Penrose charges; they  will be discussed in a forthcoming paper.

Finally, our analysis below will involve a  total derivative current with
\begin{equation}\label{eq:Z[K]}
    Z[K]^{\mu\nu\rho} = 12 K^{[\mu\nu} \Gamma\indices{^{\rho\beta]}_{|\beta}} + 4h\indices{_\beta^{[\mu}} \tilde{K}^{\nu\rho\beta]}
\end{equation}
where $K$ is a CKY tensor and $\tilde{K}$ is defined in eq.~\eqref{eq:Ktilde_Khat_def_intro}. This will be seen to arise in the relation of the ADM charges to the Penrose ones.

\section{The Penrose currents}
\label{sec:Penrose2Form}

In this section, we discuss the Penrose currents for the free graviton in $d$-dimensional Minkowski space. In particular, we investigate improvement terms that make them conserved off-shell in certain cases.

\subsection{The Penrose 2-form current away from sources}

For any 2-form $K_{\mu\nu}$ there is a 2-form 
\begin{equation}\label{eq:Penrose_2form}
    Y[K]_{\mu\nu} \equiv R_{\mu\nu\alpha\beta}K^{\alpha\beta}
\end{equation}
where $R_{\mu\nu\alpha\beta}$ is the linearised curvature tensor \eqref{eq:DefRiemann}.
We now show that the condition for this to be conserved, i.e. $\partial^\nu Y[K]_{\mu\nu} = 0$, when $R_{\mu\nu}=0$ is that $K$ is a CKY tensor. We have
\begin{equation}\label{eq:DivPenroseWorking}
    \partial^\nu Y[K]_{\mu\nu} = (\partial^\nu R_{\mu\nu\alpha\beta}) K^{\alpha\beta} + R_{\mu\nu\alpha\beta} \partial^\nu K^{\alpha\beta}
\end{equation}
From the contracted Bianchi identity 
\begin{equation}\label{eq:Contracted_Bianchi_Identity}
	\partial^\nu R_{\mu\nu\alpha\beta} =-2 \partial_{[\alpha}R_{\beta]\mu}
\end{equation}
with $R_{\mu\nu}$ the linearised Ricci tensor,
the first term on the right-hand side of eq.~\eqref{eq:DivPenroseWorking} vanishes if the vacuum Einstein equations $R_{\mu\nu}=0$ hold.
The remaining term in eq.~\eqref{eq:DivPenroseWorking} will also vanish, using $R_{\mu\nu}=0$ and the Bianchi identity $R_{\mu[\nu\alpha\beta]}=0$, provided that $K$ satisfies
\begin{equation} \label{eq:PenroseConservationCondition}
    \partial_\nu K_{\alpha\beta} = a_{\nu\alpha\beta} + \eta_{\nu[\alpha}b_{\beta]}
\end{equation}
where $a_{\nu\alpha\beta}=a_{[\nu\alpha\beta]}$ is a  3-form and $b$ is a 1-form. Indeed, we would then have
\begin{equation} \label{eq:DivPenrose}
	\partial^\nu Y[K]_{\mu\nu} = R_{\mu[\nu\alpha\beta]} a^{\nu\alpha\beta} - R_{\mu\beta}b^\beta = 0 
\end{equation}
Antisymmetrising eq.~\eqref{eq:PenroseConservationCondition} over the $\nu$, $\alpha$ and $\beta$ indices fixes $a_{\nu\alpha\beta} = \partial_{[\nu} K_{\alpha\beta]}$, while contracting the $\nu$ and $\alpha$ indices fixes $b_\beta = \frac{2}{d-1} \partial^\gamma K_{\gamma\beta}$. Hence, the condition on $K$ for $Y[K]$ to be conserved when $R_{\mu\nu}=0$ is
\begin{equation}\label{eq:CKY_equation}
    \partial_\nu K_{\alpha\beta} =\tilde   K_{\nu\alpha\beta} + 2 \eta_{\nu[\alpha} \hat K_{\beta]}
\end{equation}
where 
\begin{equation} \label{eq:Khat_Ktilde_def}
   \hat K_\mu \equiv \frac{1}{d-1}\partial^\nu K_{\nu\mu}\qc \tilde{K}_{\nu\alpha\beta}\equiv\partial_{[\nu} K_{\alpha\beta]}
\end{equation}
Eq.~\eqref{eq:CKY_equation} is the condition that $K_{\mu\nu}$ is a rank-2 CKY tensor of Minkowski space. We  refer to currents $Y[K]$ of the form eq.~\eqref{eq:Penrose_2form} with $K$ a CKY tensor as `Penrose currents' or `Penrose 2-forms'.

Given a CKY tensor $K$, we can then integrate $\star Y[K]$ over a codimension-2 cycle $\Sigma_{d-2}$ to define a conserved charge
\begin{equation}
    Q[K] \equiv \int_{\Sigma_{d-2}} \star Y[K]
\end{equation}
Since $Y[K]$ is conserved in regions where $R_{\mu\nu}=0$, the value of $Q[K]$ remains unchanged as $\Sigma_{d-2}$ is deformed through such regions.\footnote{That is, the charge is unchanged if the $(d-1)$-dimensional surface swept out by $\Sigma_{d-2}$ as it is deformed is entirely contained in a region in which $R_{\mu\nu}=0$.} 
This statement holds irrespective of the field equations. However, when the field equations $G_{\mu\nu} = T_{\mu\nu}$ hold this statement is equivalent to  saying that $Q[K]$ is conserved  in regions  where $T_{\mu\nu}=0$, i.e.\ in regions without sources.

On Minkowski space, the CKY tensors can be found explicitly. It is shown in Appendix~\ref{app:CKY_Minkowski} that taking two further derivatives of eq.~\eqref{eq:CKY_equation} leads to the integrability condition $\partial_\mu \partial_\nu \partial_\rho K_{\alpha\beta}=0$. It is then simple to demonstrate that the most general solution of eq.~\eqref{eq:CKY_equation} is \cite[eq.~(6.4.8)]{penrose_rindler_1986}
\begin{equation} \label{eq:CKY_Solution}
	K_{\alpha\beta} = \A_{\alpha\beta} + \B_{[\alpha}x_{\beta]} + \C_{\alpha\beta\gamma} x^\gamma + 2x_{[\alpha}\D_{\beta]\gamma}x^\gamma + \frac{1}{2} \D_{\alpha\beta} x_\gamma x^\gamma
\end{equation}
where $\A$, $\B$, $\C$, and $\D$ are constant antisymmetric tensors.
Therefore, in $d$ dimensions there are $d(d+1)(d+2)/6$ rank-2 CKY tensors. 
Particularly, in $d=4$ dimensions, there are 20 independent solutions. 

An important result for our analysis is that the divergence of a rank-2 CKY tensor on Minkowski space gives a Killing vector\footnote{For CKY tensors on a curved space, it is not guaranteed that the covariant divergence $\hat{K}_\mu = (d-1)^{-1} \nabla^\nu K_{\nu\mu}$ is a Killing vector but $G_{\mu\nu}\hat{K}^\nu$ remains covariantly conserved --- that is, $\nabla^\mu(G_{\mu\nu}\hat{K}^\nu)=0$ --- as a result of the integrability condition $(d-2)\nabla_{(\mu}\hat K_{\nu)} = R\indices{^\rho_{(\mu}} K_{\nu)\rho}$ satisfied by CKY tensors \cite{Lindstrom2022Killing-YanoCurrents}.} $\hat{K}_\alpha $ and the exterior derivative gives a closed CKY tensor $\tilde{K}_{\alpha\beta\gamma}$. We now show this.
Explicitly, taking the divergence of the general rank-2 CKY tensor in eq.~\eqref{eq:CKY_Solution} gives
\begin{equation}\label{eq:Khat_solution}
    \hat{K}_\alpha = \frac{1}{d-1} \partial^\beta K_{\beta\alpha} = -\frac{1}{2} \B_\alpha + \D_{\alpha\beta}x^\beta 
\end{equation}
which is the form of the Killing vectors of Minkowski space.
The $\hat{K}$'s for the $\B$-type CKY tensors are the translational Killing vectors, while the $\hat{K}$'s for the $\D$-type CKY tensors are the Killing vectors for Lorentz transformations.
We see that only CKY tensors of the $\B$- and $\D$-types in eq.~\eqref{eq:CKY_Solution} correspond to Killing vectors as the $\A$- and $\C$-type terms are divergenceless. In section~\ref{sec:PenroseElectricRelation}, we will relate the Penrose currents for $\B$- and $\D$-type CKY tensors to the ADM currents for the corresponding Killing vectors $\hat{K}_\mu$.

A Killing-Yano (KY) 2-tensor is one that satisfies
\begin{equation}\label{eq:KY_equation}
    \partial_\nu K_{\alpha\beta} = \partial_{[\nu} K_{\alpha\beta]} 
\end{equation}
so that it is a CKY tensor whose divergence $\hat{K}_\mu=0$.
The general KY tensor is then $K_{\alpha\beta}=f_{\alpha\beta}$ where
\begin{equation} \label{eq:KY_Solution}
f_{\alpha\beta}	
  = \A_{\alpha\beta} + \C_{\alpha\beta\gamma} x^\gamma 
\end{equation}
In Minkowski space, this is co-exact: it can be written explicitly as the divergence of a  3-form
\begin{equation}\label{eq:f=divF}
    f_{\alpha\beta} = \partial^\gamma \mathcal{F}_{\alpha\beta\gamma} 
\end{equation}
where
\begin{equation}\label{eq:rank3KYPotential}
    \mathcal{F}_{\alpha\beta\gamma} = \frac{3}{d-2} \A_{[\alpha\beta}x_{\gamma]} + \frac{1}{2(2-d)} \left( \C_{\alpha\beta\gamma} x^2 - 6 x_{[\alpha}\C_{\beta\gamma]\sigma}x^\sigma \right)
\end{equation}
(We can explicitly verify that $\mathcal{F}$ in fact satisfies the rank-3 CKY equation, given in Appendix~\ref{app:AppendixCKY}, on Minkowski space.)
In section~\ref{sec:KaluzaKlein}, we will consider the case in which Minkowski space is replaced with $\mathbb{R}^{1,d-1-n} \times T^n$ where the KY tensors need not be co-exact.

The exterior derivative of the general rank-2 CKY tensor in eq.~\eqref{eq:CKY_Solution} gives
\begin{equation}\label{eq:Ktilde_solution}
  \tilde{K}_{\alpha\beta\gamma} = \partial_{[\alpha}K_{\beta\gamma]} = \C_{\alpha\beta\gamma} + 3\D_{[\alpha\beta} x_{\gamma]}
\end{equation}
which is precisely the form of a general rank-3 closed CKY tensor (see Appendix~\ref{app:CKY_Minkowski} for the definition of CKY tensors of general rank).
We note that only the CKY tensors of the $\C$- and $\D$-types in eq.~\eqref{eq:CKY_Solution} contribute to rank-3 closed CKY tensors $\tilde{K}_{\mu\nu\rho}$.
The rank-2 CKY $K$ is itself closed if $\tilde{K}_{\mu\alpha\beta} = \partial_{[\mu}K_{\alpha\beta]}$ vanishes. So the general closed CKY tensor is $K_{\alpha\beta} = \sigma_{\alpha\beta}$ where
\begin{equation} \label{eq:CCKY_Solution}
	\sigma_{\alpha\beta} = \A_{\alpha\beta} + \B_{[\alpha}x_{\beta]} 
\end{equation}
On Minkowski space, closed CKY tensors are in fact exact,
\begin{equation} \label{eq:CCKY_Solutionexact}
	\sigma_{\alpha\beta} = \partial_{[\alpha} \Sigma_{\beta]}
\end{equation}
where
\begin{equation}
    \Sigma_\alpha = -\A_{\alpha\beta} x^\beta -\frac{1}{2} \left( \frac{1}{2} \B_\alpha x^2 - \B_\beta x_\alpha x^\beta \right)
\end{equation}
We can then verify that $\Sigma_\alpha$ satisfies the rank-1 CKY equation; that is, $\Sigma_\alpha$ is a conformal Killing vector.

One important property of CKY tensors which we will use throughout is that the Hodge dual of a rank-$p$ CKY tensor is, itself, a rank-$(d-p)$ CKY tensor (see Appendix~\ref{app:AppendixCKY}). In particular, the dual of a KY tensor (i.e. a tensor of the form of eq.~\eqref{eq:KY_Solution}) is a closed CKY tensor (i.e. of the form of eq.~\eqref{eq:CCKY_Solution}) and vice-versa. This property is true of CKY tensors on any manifold, and several of the other properties discussed above are also true on more general spaces \cite{Tachibana1969OnSpace, Kashiwada1968, Frolov2008HigherdimensionalVariables, Howe2018SCKYT, Lindstrom2022Geometrycurrents}.

\subsection{Improved Penrose 2-form in the presence of sources}
\label{sec:PenroseImprovement}

The Penrose 2-form \eqref{eq:Penrose_2form} is conserved provided that the Ricci tensor vanishes.
We now consider adding sources and suppose Einstein's equation $G_{\mu\nu}=T_{\mu\nu}
$ is satisfied for some conserved energy-momentum tensor $T_{\mu\nu}$.
As has been seen, the Penrose 2-form $Y[K]$ is conserved in the region in which $T_{\mu\nu}=0$.

However, there exists an improvement which is conserved without use of the field equations for KY tensors. We define the `improved Penrose 2-form'
\begin{equation} \label{eq:Def_ImprovedPenrose}
	Y_+[K]_{\mu\nu} \equiv R_{\mu\nu\alpha\beta} K^{\alpha\beta} + 4 R\indices{^\alpha_{[\mu}} K_{\nu]\alpha} + R K_{\mu\nu}
\end{equation}
In regions without sources, this reduces to the Penrose 2-form $Y[K]$.
Now, using eq.~\eqref{eq:Contracted_Bianchi_Identity} as well as the contracted Bianchi identity
\begin{equation} \label{eq:Contracted_Bianchi_Identity2}
    2\partial^\alpha R_{\alpha\beta} = \partial_\beta R
\end{equation}
we find 
\begin{equation} \label{eq:DivImprovedPenrose}
	\partial^\nu Y_+[K]_{\mu\nu} = 2(d-3) G_{\mu\alpha} \hat K ^\alpha 
\end{equation}
This is proportional to the conserved current \eqref{eq:jh[k]} with Killing vector $k^\alpha = 2(d-3) \hat{K}^\alpha$.
If $K$ is a KY tensor (i.e. $\hat{K} =0$), $Y_+[K]$ is conserved and is precisely the current of Kastor and Traschen \cite{Kastor2004ConservedTensors}. 
In what follows, we will use the improved 2-form current $Y_+[K]$ for general CKY tensors. 

We may now define a quantity by integrating $\star Y_+[K]$ over a codimension-2 cycle:
\begin{equation}
    Q_+[K] \equiv \int_{\Sigma_{d-2}} \star Y_+[K]
\end{equation}

If $K$ is KY tensor (i.e. an $\A$- or $\C$-type CKY tensor), then $Y_+[K]$ is conserved off-shell, i.e. without using the field equations or any condition on $T_{\mu \nu}$, at all points and so $Q_+[K]$ is conserved. This means that
the value of $Q_+[K]$ is   unchanged as $\Sigma_{d-2}$ is deformed arbitrarily (including deformations  through regions where $T_{\mu\nu}\neq0$).
For the $\B$- and $\D$-type CKY tensors, however, $Y_+[K]$ is only conserved at points at which
$R_{\mu\nu}=0$, so that on-shell this means at points at which $T_{\mu\nu}=0$.
Then for surfaces $\Sigma_{d-2} $ that lie in a region $\mathcal{R}$ in which $R_{\mu\nu}=0$, the charge $Q_+[K]$ given by the integral of $\star Y_+[K]$ over $\Sigma_{d-2}$ is unchanged under deformations that keep the surface in the region $\mathcal{R}$. In this region 
$Y_+[K]=Y[K]$, so  
\begin{equation}
    Q_+[K] = Q[K] \qquad \text{when $\Sigma_{d-2} \subset \mathcal{R}$}
\end{equation}
In this paper, we will primarily restrict ourselves to the case where $\Sigma_{d-2} \subset \mathcal{R}$ for some region 
$\mathcal{R}$ in which $R_{\mu\nu}=0$.

For general CKY tensors $K$, the proportionality of $\partial^\nu Y_+[K]_{\mu\nu}$ and the 1-form current $j[k]_\mu = G_{\mu\alpha}k^\alpha$, suggests a link between the improved Penrose 2-form $Y_+[K]$ and the secondary current $J[k]$. We will discuss this in detail in the following sections. 
The improvement terms in eq.~\eqref{eq:Def_ImprovedPenrose} are not unique and there are other improvements of the Penrose 2-form which reduce to eq.~\eqref{eq:Penrose_2form} on-shell (c.f. \cite{Lindstrom2021NewTensors}). The particular combination  in eq.~\eqref{eq:Def_ImprovedPenrose} is chosen to simplify the relation with the 1-form current $j[k]$. 

\subsection{Triviality of Penrose charges for KY tensors in $d>4$}
\label{trivial_KY_charges}

In this section we show that the Penrose charges for KY tensors in Minkowski space  vanish for $d>4$. 
In Ref.~\cite{Benedetti2023GeneralizedGravitons} it was shown that
$\star Y[\A]$ and $\star Y[\C]$ are exact $(d-2)$-forms in $d>4$ when $R_{\mu\nu}=0$, and it was concluded that $Q[\A]$ and $Q[\C]$ vanish. Here we improve on their argument, showing that 
$\star Y_+[\A]$ and $\star Y_+[\C]$ are exact $(d-2)$-forms \emph{off-shell}, i.e. without requiring $R_{\mu\nu}=0$, and conclude that the corresponding Penrose charges $Q_+[\A]$ and $Q_+[\C]$ vanish identically, and this remains true even when the surface of integration is deformed through regions in which $R_{\mu\nu}\neq 0$.
Note that these results are for Minkowski space and there are modifications when considering toroidal compactifications.\footnote{
In sections~\ref{sec:KaluzaKlein} and~\ref{sec:examples}
we study linearised gravity on the product of Minkowski space with a torus and find, for example, that they can have non-zero $\C$-type Penrose charges.}

For a KY tensor $K_{\mu\nu}=f_{\mu\nu}$ given by eq.~\eqref{eq:KY_Solution}, the corresponding improved Penrose 2-form $Y _+[f]$ is the divergence of a 3-form built from the linearised curvature tensor and the rank-3 CKY tensor $\mathcal{F}$ defined in eq.~\eqref{eq:rank3KYPotential}; that is, we have
\begin{equation} \label{eq:ImprovedPenrosePotential}
    Y_+[f]_{\mu\nu} =  \frac{d-2}{d-4}\,\partial^\rho \mathscr{Y}_{\mu\nu\rho}
\end{equation}
for $d>4$, where $\mathscr{Y}$ is the 3-form with components
\begin{equation} \label{eq:Xisw}
   \mathscr{Y}_{\mu\nu \rho} = 3 R\indices{_{[\mu\nu}^{\alpha\beta}} \mathcal{F}\indices{_{\rho]\alpha\beta}} -6 R\indices{^\beta_{[\mu}} \mathcal{F}\indices{_{\nu\rho]\beta}} + R \mathcal{F}_{\mu\nu\rho}
\end{equation}
This follows from  eq.~\eqref{eq:f=divF} and the contracted Bianchi identities \eqref{eq:Contracted_Bianchi_Identity} and \eqref{eq:Contracted_Bianchi_Identity2}, as well as $R_{[\mu\nu\rho]\sigma}=0$ but does not require assuming $R_{\mu\nu}=0$. 
Since the right-hand side of eq.~\eqref{eq:Xisw} depends on $h_{\mu\nu}$ only through the linearised curvature tensor, \emph{it is globally defined irrespective of whether $h_{\mu\nu}$ is globally defined or not} as the curvature tensor is globally defined. 
Therefore, in $d>4$ Minkowski space, integrals of $\star Y_+[f]$ over codimension-2 cycles vanish  by Stokes' theorem. Then the only non-zero Penrose charges in $d>4$ Minkowski space come from the CKY tensors of the $\B$ and $\D$ types (that is, those which are not KY tensors). These are precisely the ones which correspond to non-zero Killing vectors $\hat{K}$ in eq.~\eqref{eq:Khat_solution}.

Note that in $d=4$, eq.~\eqref{eq:ImprovedPenrosePotential} is no longer valid so the improved Penrose current for KY tensors cannot be written as the divergence of a tensorial 3-form and the associated Penrose charges can be non-vanishing.

While eq.~\eqref{eq:ImprovedPenrosePotential} is valid without using the field equations, in regions away where $R_{\mu\nu}=0$, the left-hand side of eq.~\eqref{eq:ImprovedPenrosePotential} reduces to the Penrose 2-form $Y[f]$ and the final two terms of $\mathscr{Y}$ in eq.~\eqref{eq:Xisw} vanish. For this case, Ref.~\cite{Benedetti2023GeneralizedGravitons} has  given a similar covariant 3-form potential for the Penrose 2-forms associated with KY tensors.

We note that Kastor and Traschen \cite{Kastor2004ConservedTensors} have also given a 3-form potential for the improved Penrose current when $K$ is a KY tensor (which they refer to as a Yano current). Theirs, however, is not a covariant expression in terms of $h$. Therefore, if we allow non-global gauge-field configurations (as we are here) then we must re-examine the contributions of total derivative terms. Our 3-form potential in eq.~\eqref{eq:Xisw}, in contrast, is covariant. As is necessary from the Poincar\'{e} lemma, these two 3-form potentials differ by the divergence of a 4-form. 
Furthermore, Kastor and Traschen consider the integration of the Penrose current for a KY tensor over a $(d-2)$-dimensional space $\Sigma_{d-2}$ with a $(d-3)$-dimensional boundary $\partial \Sigma_{d-2}$. In this case the charge $\int_{\Sigma_{d-2}}\star Y_+[f]$ can be written as an integral over the boundary 
\begin{equation}
    \int_{\Sigma_{d-2}}\star Y_+[f] = \frac{d-2}{d-4} \int_{\partial \Sigma_{d-2}} \star \mathscr{Y}
\end{equation}
for $d>4$. 
For surfaces $\Sigma_{d-2}$ that are closed this clearly vanishes, and we only consider closed surfaces in this paper.

\section{Relation between Penrose charges and ADM charges}
\label{sec:PenADM}

In this section we analyse the connection between the charges constructed from the improved Penrose 2-form $Y_+[K]_{\mu\nu}$ and the ADM charges associated with the Killing vectors of Minkowski space $k_\mu$.

We have seen in section~\ref{sec:Penrose2Form} that the divergence of a CKY tensor on flat space is a Killing vector.
It will be convenient to define
\begin{equation}\label{eq:k=2(d-3)Khat}
    k_\mu = 2(d-3) \hat{K}_\mu
\end{equation}
Then eq.~\eqref{eq:DivImprovedPenrose} gives a relation between the improved Penrose 2-form and the 1-form primary current $j[k]_\mu = G_{\mu\nu}k^\nu$  which can be written as
\begin{equation}\label{eq:DivImprovedPenrose_j[k]}
    \partial^\nu Y_+[K]_{\mu\nu} = j[k]_\mu 
\end{equation}
The secondary current $J[k]$ given in eq.~\eqref{eq:J[k]_def} satisfies $\partial^\nu J[k]_{\mu\nu}=j[k]_\mu$ so that
\begin{equation}\label{eq:Div}
    \partial^\nu Y_+[K]_{\mu\nu} = \partial^\nu J[k]_{\mu\nu} 
\end{equation}

It follows from eq.~\eqref{eq:Div} and the Poincar\'{e} lemma that, on contractible open sets, $Y_+[K]$ and $J[k]$ should be related by the divergence of some locally defined 3-form $Z$. Moreover, from the algebraic Poincar\'{e} lemma it is to be expected that $Z$ has a local expression in terms of the graviton field and $K$.
Indeed, we show in appendix~\ref{app:AppendixRiemann} that
\begin{align}\label{eq:ADM_ImprovedPenrose_Relation}
    Y_+[K]^{\mu\nu} = J[k]^{\mu\nu} + \partial_\rho Z[K]^{\mu\nu\rho} 
\end{align}
where
\begin{equation}\label{eq:UpsilonGeneral}
    Z[K]^{\mu\nu\rho} = 12 K^{[\mu\nu} \Gamma\indices{^{\rho\beta]}_{|\beta}} + 4h\indices{_\beta^{[\mu}} \tilde{K}^{\nu\rho\beta]}
\end{equation}
with $\tilde{K}$ given in eq.~\eqref{eq:Khat_Ktilde_def}.
We emphasise that $Z[K]$ is a 3-form which depends explicitly on the gauge field $h_{\mu\nu}$.
The divergence of eq.~\eqref{eq:ADM_ImprovedPenrose_Relation} then gives eq.~\eqref{eq:DivImprovedPenrose_j[k]}. 

Integrating eq.~\eqref{eq:ADM_ImprovedPenrose_Relation} over a codimension-2 cycle $\Sigma_{d-2}$ gives
\begin{equation} \label{eq:ADM_Electric_Charges}
    Q[K] = Q[k] + \int_{\Sigma_{d-2}} \dd{\star Z[K]} 
\end{equation}
We shall generally restrict ourselves to surfaces $\Sigma_{d-2}$ contained in a region where $R_{\mu\nu}=0$, i.e. in a region without sources if the field equations hold.
Recall that $Y_+[K]$ and $J[k]$ are only conserved in such regions since $\partial_\nu J[k]^{\mu\nu} = j[k]^\mu = T^{\mu\nu}k_\nu$. Therefore, the integral is unchanged under any deformation of the surface $\Sigma_{d-2}$ that does not cross a region where $R_{\mu\nu}\neq 0$. 

If $h$ is a globally defined tensor, then $Z$ is also globally defined, so that $\int\dd\star Z[K]=0$ and eq.~\eqref{eq:ADM_Electric_Charges} reduces to
\begin{equation} \label{eq:ADM_Electric_Charges2}
    Q[K] = Q[k] 
\end{equation}
Then $Q[K]$ depends only on $\hat K$ and is precisely the ADM charge for the Killing vector $k= 2(d-3) \hat{K}$.

The general form of $\hat K$ is given in eq.~\eqref{eq:CKY_Solution} in terms of constant antisymmetric tensors $\A$, $\B$, $\C$, and $\D$.
We will use the notation $Q[\A]$, $Q[\B]$, $Q[\C]$ and $Q[\D]$ to refer to the Penrose charges $Q[K]$ with the CKY tensor $K$ given by the relevant term in eq.~\eqref{eq:CKY_Solution}.
Then if $h$ is a globally defined tensor we have
\begin{equation} 
    Q[\A] = Q[\C] = 0
\end{equation}
as the $\A$- and $\C$-type CKY tensors do not contribute to $\hat{K}$.
From subsection \ref{trivial_KY_charges} we already knew this to be the case for $d>4$, so here we learn that this also applies for $d=4$ if $h$ is globally defined.
In four dimensions, we have
\begin{equation} \label{eq:B_and_D_type_ADM_link}
    Q[\B] = -\B_\mu P^\mu, \qquad
    Q[\D] = \D_{\mu\nu} L^{\mu\nu}
\end{equation}
where $P^\mu,L^{\mu\nu}$ are the ADM momentum and angular momentum defined in eq.~\eqref{eq:ADM_charge_definition}.

The current $Y_+[K]^{\mu\nu}$ is invariant under the gauge transformation \eqref{eq:h_gauge_transformation} while $J[k]^{\mu\nu}$ is not.
Under the gauge transformation \eqref{eq:h_gauge_transformation}, $Z[K]^{\mu\nu\rho} $ changes by
\begin{equation}
    \delta Z[K]^{\mu\nu\rho} = -8\tilde{K}^{[\mu\nu\rho} \partial_\sigma \zeta^{\sigma]} + 6(d-3) \hat{K}^{[\mu} \partial^\nu \zeta^{\rho]}
\end{equation}
so that it follows from eq.~\eqref{eq:ADM_ImprovedPenrose_Relation} that
\begin{equation}\label{eq:J[k]_gauge_transf}
    \delta J[k]^{\mu\nu} = \partial_\rho \left( 8\tilde{K}^{[\mu\nu\rho} \partial_\sigma \zeta^{\sigma]} - 6(d-3) \hat{K}^{[\mu} \partial^\nu \zeta^{\rho]} \right)
\end{equation}
as can be explicitly verified.
If $h$ is a globally defined tensor then the gauge transformation is \eqref{eq:h_gauge_transformation} with $\zeta$ a globally defined 1-form. Then $\delta J[k]$ in eq.~\eqref{eq:J[k]_gauge_transf} is the total derivative of a globally defined 3-form and it follows that the integral of this variation vanishes so that the charge $Q[k]$ is gauge invariant.

We now turn to the case in which $h_{\mu\nu}$ is not a globally defined field configuration, either with a Dirac string singularity or defined in patches with non-trivial transition functions involving the gauge transformation \eqref{eq:h_gauge_transformation}.
Then $Z[K]$ is not globally defined in general and the total derivative term $\int  \dd{\star Z[K]}$ in \eqref{eq:ADM_Electric_Charges} need not vanish but is instead a topological term.
Moreover the variation $\delta J[k]^{\mu\nu}$ under a gauge transformation given in eq.~\eqref{eq:J[k]_gauge_transf} is not globally defined in general\footnote{If $h$ is defined in patches then we can consider different gauge transformations in each patch, so that there could be a different $\zeta$ in each patch, and as a result $\zeta$ need not be a globally defined 1-form in general.}
so that the charge $Q[k]$ defined in eq.~\eqref{qkisb} is no longer gauge-invariant. Then the definition of ADM charges needs modification for non-globally defined $h_{\mu\nu}$.
Adding the topological term to the ADM charges $Q[k]$ as in eq.~\eqref{eq:ADM_Electric_Charges} gives a gauge-invariant result, and provides such a covariantisation.
The results of subsection \ref{trivial_KY_charges} place further restrictions on the topological charges for $d>4$, which we discuss in the next section.
In section~\ref{sec:four_dimensions} we discuss the case $d=4$.

\section{Analysis of Penrose charges in $d>4$}
\label{sec:PenroseElectricRelation}

We have seen in the last section that for globally defined $h_{\mu\nu}$ the Penrose charges give the  ADM charges, while if 
$h_{\mu\nu}$ is not globally defined, then the Penrose charges give the naive ADM charges \eqref{qkisb} plus a topological charge.
In this section, we further investigate the general case in which 
$h_{\mu\nu}$ need not be not globally defined for dimensions $d>4$.

In dimensions $d>4$, the results of subsection \ref{trivial_KY_charges} restrict the charges.
The right-hand-side of eq.~\eqref{eq:ImprovedPenrosePotential} is a globally defined total derivative even if $h_{\mu\nu}$ is not globally defined, so the CKY tensors with coefficients $\A$ or $\C$ in eq.~\eqref{eq:CKY_Solution} give a current
$Y_+[K]_{\mu\nu} $ that is co-exact and so give zero charge
\begin{equation} 
    Q[\A] = Q[\C]=0
\end{equation}

Consider next the $\B$-type CKY tensors, $K_{\mu\nu} = \B_{[\mu}x_{\nu]}$. From eq.~\eqref{eq:Khat_solution}, these correspond to constant  translational Killing vectors $\hat{K}_\mu = - \frac{1}{2} \B_\mu$ and give $\tilde{K}_{\mu\nu\rho}=0$. 
Then
\begin{equation}\label{eq:UpsilonB}
    Z[\B]^{\mu\nu\rho} = 12 \B^{[\mu}x^{\nu} \Gamma\indices{^{\rho\beta]}_{|\beta}} 
\end{equation}
and
\begin{align}\label{eq:ADM_ImprovedPenrose_RelationB}
    Y_+[\B]^{\mu\nu} = J[k]^{\mu\nu} + \partial_\rho Z[\B]^{\mu\nu\rho} 
\end{align}
Neither of the two terms on the right hand side are gauge-invariant in general but their sum is, so that 
adding an identically conserved term $\dd^\dagger Z$ to $J$ gives a gauge-invariant current.
Thus $Y_+[\B]^{\mu\nu}$ can be viewed as an \emph{improved} ADM current and the integral of this gives an improved ADM charge
\begin{equation}\label{eq:ADM_Penrose_B}
     Q[\B] = Q[k] + \int_{\Sigma_{d-2}} \dd\star Z[\B] 
\end{equation}
with $\hat{K}_\mu = \frac{1}{2(d-3)} k_\mu = -\frac{1}{2}\B_\mu$. This is unchanged under deformations of $\Sigma_{d-2}$ which do not cross regions where $T_{\mu\nu}\neq0$.
This agrees with the usual ADM charge when $h$ is globally defined and is gauge-invariant even when $h$ is not globally defined, so that it provides a natural improved definition of the ADM charge, which we denote $\mathbb{P}^\mu$, for a constant Killing vector $k^\mu$.
It can be written as
\begin{equation}
  \mathbb{P}^\mu k_\mu = -\frac{1}{2(d-3)} \int_{\Sigma_{d-2}} R_{\mu\nu\alpha\beta} k^\alpha x^\beta \dd{\Sigma^{\mu \nu}}
\end{equation}

The Penrose 2-form $Y[K]$ for the $\B$-type CKY tensors can be written
\begin{equation}
    Y[\B]_{\mu\nu} = \partial_{[\mu} b_{\nu]} 
\end{equation}
where $b$ is the 1-form 
\begin{equation}\label{eq:b_1form}
    b_\nu = 2\Gamma\indices{^{\alpha\beta}_{|\nu}} \B_\alpha x_\beta + h_{\nu\alpha} \B^\alpha
\end{equation}
Under a gauge transformation \eqref{eq:h_gauge_transformation} this transforms as a 1-form gauge field
\begin{equation}
   \delta  b_\mu = \partial _\mu \chi
\end{equation}
where
\begin{equation}
    \chi = 2\partial^{[\alpha} \zeta^{\beta]} \B_\alpha x_\beta + 3\zeta^\alpha \B_\alpha
\end{equation}
Since we assume that $\Sigma_{d-2}$ is contained in a region where $R_{\mu\nu}=0$, we have $Y[K] = Y_+[K]$ on $\Sigma_{d-2}$, so the Penrose charge for the $\B$-type CKY tensors is
\begin{equation}
    Q[\B] = \int_{\Sigma_{d-2}} \star \dd{b}
\end{equation}
which is  the electric charge for the 1-form gauge field $b_\mu$.
 
Now, consider the $\D$-type CKY tensors, 
$$K_{\mu\nu} = 2x_{[\mu}\D_{\nu]\rho}x^\rho + \frac{1}{2}\D_{\mu\nu}x^2$$
These have both $\hat{K}$ and $\tilde{K}$ non-zero:
\begin{equation}
    \hat{K}_\mu = \frac{1}{2(d-3)} k_\mu = \D_{\mu\nu} x^\nu \qc \tilde{K}_{\mu\nu\rho} = 3\D_{[\mu\nu}x_{\rho]} 
\end{equation}
from eqs.~\eqref{eq:Khat_solution} and \eqref{eq:Ktilde_solution}, so that $k$ is a Killing vector generating a Lorentz transformation. Then
\begin{align}\label{eq:ADM_ImprovedPenrose_RelationD}
    Y_+[\D]^{\mu\nu} = J[k]^{\mu\nu} + \partial_\rho Z[\D]^{\mu\nu\rho} 
\end{align}
As before, adding the identically conserved improvement term $\dd^\dagger Z$ to $J[k]$ gives a gauge-invariant current.
Integrating this gives a covariant improved ADM angular momentum which we denote $\mathbb{L}^{\mu\nu}$,
\begin{equation}\label{eq:ADMcov}
    \frac{1}{2} \Lambda_{\mu\nu} \mathbb{L}^{\mu\nu} = \frac{1}{4(d-3)} \int_{\Sigma_{d-2}} R_{\mu\nu\alpha\beta} \, K^{\alpha \beta} \dd{\Sigma^{\mu \nu}}
\end{equation}
with
\begin{equation}
    K_{\mu\nu } = 2 x_{[\mu} \Lambda_{\nu]\rho} x^\rho + \frac{1}{2}\Lambda_{\mu\nu}x^2
\end{equation}
This charge agrees with the ADM angular momentum when $h$ is globally defined and is gauge invariant even when $h$ is not globally defined.

Note that in this case eq.~\eqref{eq:ADM_Electric_Charges} can also be written as 
\begin{equation}\label{eq:Penrose_Dtype}
    Q[\D] = Q[k] + Q[\lambda] + \int_{\Sigma_{d-2}} \dd\star W[\D] 
\end{equation}
where $Q[\lambda]$ is given by eq.~\eqref{qkisl} with $\tilde{\lambda}_{\mu\nu\rho}=(-1)^{d-1}(d-1) \tilde{K}_{\mu\nu\rho}$ and $W[\D]$ is given by 
\begin{equation}\label{eq:W[K]}
    W[K]^{\mu\nu\rho} = 12 K^{[\mu\nu} \Gamma\indices{^{\rho\beta]}_{|\beta}}
\end{equation}
with $K$ a $\D$-type CKY tensor.
The variation of the total derivative term involving $W[\D]$ cancels the gauge variations of both $Q[k]$ and $Q[\lambda]$. However, there seems to be no gauge-invariant way of separating the total charge $\int\dd\star W[\D]$ into a covariantisation of $Q[k]$ plus a covariantisation of $Q[\lambda]$.
We will return to this point in a future publication, where we will show that this term can alternatively be understood as the covariantisation of one of the magnetic charges discussed in Ref.~\cite{HullYetAppear} that arises as an electric charge for the dual graviton.

\section{Dual charges in $d>4$}
\label{sec:d-3_form_charges}

For each $K$, the Penrose 2-form current $Y[K]$ is conserved on-shell and can be integrated over a $(d-2)$-cycle to give a conserved charge $Q[K]$. It is a topological charge that generates a 1-form symmetry \cite{Gaiotto2015GeneralizedSymmetries}.
The dual $(d-2)$-form current $\star Y[K]$ is also conserved if $K$ is chosen so that $Y[K]$ is closed. If this is the case, it can  be integrated over a 2-cycle to give a conserved dual charge $q[K]$. This dual  charge then generates a $(d-3)$-form symmetry \cite{Gaiotto2015GeneralizedSymmetries}.

We now consider the conditions on $K$ for the closure of the Penrose 2-form. 
Taking the curl of eq.~\eqref{eq:Penrose_2form} and using the CKY equation \eqref{eq:CKY_equation}, we find
\begin{equation}
\begin{split}\label{eq:dY_working}
    \partial_{[\rho} Y[K]_{\mu\nu]} &= R\indices{_{\alpha\beta[\mu\nu}} \tilde{K}\indices{_{\rho]}^{\alpha\beta}} + 2 R\indices{_{\alpha\beta[\mu\nu}} \delta_{\rho]}^\alpha \hat{K}^\beta \\
    &= R\indices{_{\alpha\beta[\mu\nu}} \tilde{K}\indices{_{\rho]}^{\alpha\beta}}
\end{split}
\end{equation}
where we have used the differential Bianchi identity $\partial_{[\rho} R_{\mu\nu]\alpha\beta}=0$ in the first equality and the algebraic Bianchi identity $R_{\alpha[\mu\nu\rho]}=0$ in the second.
It is clear that this vanishes for closed CKY tensors $K$, for which $\tilde{K}=0$. 
For $d>4$, $Y[K]$ is closed \emph{only} when $K$ is a closed CKY tensor. However, in $d=4$ we will see in section~\ref{sec:no_dual_charges_4d} that $Y[K]$ is closed for all CKY tensors in regions where $R_{\mu\nu}=0$.
We note in passing that (in $d>4$) the improved Penrose 2-form $Y_+[K]$ is not closed unless $K$ is closed and $R_{\mu\nu}=0$, in which case it is simply equal to the Penrose 2-form $Y[K]$.
The closed CKY tensors are given in eq.~\eqref{eq:CCKY_Solution}: they are the $\A$- and $\B$-type CKY tensors.

We can form conserved charges by integrating $Y[K]$ over a 2-cycle $\Sigma_2$ which is contained in a region where $R_{\mu\nu}=0$, where $K_{\mu\nu} = \sigma_{\mu\nu}$ is a closed CKY tensor,
\begin{equation}\label{eq:dual_Penrose_charge}
    q[\sigma] \equiv \int_{\Sigma_2} Y[\sigma]
\end{equation}
The charge $q[\A]$ for the $\A$-type closed CKY tensors can be written as
\begin{equation}\label{eq:q[A]_mag_charge}
    q[\A] = \int_{\Sigma_2} \dd a
\end{equation}
where 
\begin{equation}\label{eq:a_1form}
    a_\nu = 2\Gamma\indices{^{\alpha\beta}_{|\nu}} \A_{\alpha\beta}
\end{equation}
This is the magnetic charge for the 1-form gauge field $a$ and is non-zero only when $h$ is  not a globally defined gauge field configuration.

For the $\B$-type closed CKY tensors, the charge can be written
\begin{equation}\label{eq:BtypeHolonomy}
    q[\B] = \int_{\Sigma_2} \dd b .
\end{equation}
which is the magnetic charge for the 1-form $b$ defined in eq.~\eqref{eq:b_1form}.
These charges also vanish unless the gauge field is not globally defined.

There are $d(d+1)/2$ such dual charges $q[\A]$, $q[\B]$ in $d$ dimensions. There are $d(d+1)(d+2)/6$ CKY tensors but,  as discussed in section~\ref{trivial_KY_charges}, on Minkowski spacetime the Penrose charges are trivial when $K$ is a KY tensor, so that only $Q[\B]$ and $Q[\D]$ are non-trivial.
Then the number of non-trivial Penrose charges is $d(d+1)/2$ also (for $d>4$). This is in accordance with the discussion of Ref.~\cite{Benedetti2023GeneralizedGravitons}, which argued that the equality between the number of 1-form and $(d-3)$-form symmetries was to be expected for higher-form symmetries that are charged under continuous spacetime symmetries and that this required
$Q[\A]$ and $Q[\C]$ to be trivial.
For example, in the present case, the 1-form symmetries generated by the $Q[K]$ transform non-trivially under the Lorentz group as the CKY tensors $K$ carry Lorentz indices \cite{BenedettiNoether}.
This stems from the principle that higher-form symmetries always come in dual pairs \cite{CasiniCompleteness}.
It is natural to regard the $d$ charges $Q[\B]$ as pairing with the $d$ charges $q[\B]$ and the $d(d-1)/2$ charges $Q[\D]$ as pairing with the $d(d-1)/2$ charges $q[\A]$.

It will be seen in section~\ref{sec:KaluzaKlein} that for spacetimes that are the product of Minkowski space with a torus the classification of charges and dual charges is slightly different, but the equality between the numbers of  1-form symmetries and $(d-3)$-form symmetries remains.

\section{Penrose charges in $d=4$}
\label{sec:four_dimensions}

The case of $d=4$ is special due to several properties of the Riemann tensor and the CKY tensors.

\subsection{No independent dual charges in $d=4$}
\label{sec:no_dual_charges_4d}

We have seen that for any CKY tensor $K$ the Penrose current $Y[K]$ is conserved in regions where $R_{\mu\nu}=0$, i.e. on-shell in regions away from sources. However, in $d=4$, $Y[K]$ is also closed on-shell in these regions. 
This follows from eq.~\eqref{eq:dY_working} which, in $d=4$, implies
\begin{equation}\label{eq:dY_working_4d}
    \partial_{[\rho} Y[K]_{\mu\nu]} = \frac{2}{3} \epsilon_{\mu\nu\rho\gamma} G\indices{^\gamma_\delta} (\star\tilde{K})^\delta 
\end{equation}
This result is derived in Appendix~\ref{app:closure_Y}.
The right-hand side of eq.~\eqref{eq:dY_working_4d} vanishes by the field equations in regions without sources.
Note that this result holds only in four dimensions. However, in regions where sources are present, $Y[K]$ is only closed when $K$ is a closed CKY tensor.

As $Y[K]$ is both closed and co-closed in the absence of sources, we can build charges by integrating $Y[K]$ or $\star Y[K]$ over 2-cycles.
The integral $\int \star Y[K]$ gives the Penrose charges while $\int  Y[K]$ gives further conserved charges.
However, in four dimensions, these charges are not independent of the Penrose charges. This follows from the duality of the CKY tensors in four dimensions, as we now show.

First, we recall from section~\ref{sec:Penrose2Form} that the Hodge dual of a CKY tensor is also a CKY tensor. So in four dimensions the Hodge dual of a rank-2 CKY tensor $K$ is another rank-2 CKY tensor $\star K$, giving a conserved current $Y[\star K]$.
In four dimensions there is a 20-dimensional vector space of CKY 2-tensors $K$ and the dual tensors $\star K$ form the same 20-dimensional space; Hodge duality is an automorphism of this space. The set of currents $Y[\star K]$ is precisely the same as the set of currents $Y[K]$.

Next, the closure of $Y[K]$ is equivalent to the co-closure of its Hodge dual, 
\begin{equation}
	\star{Y}[K]_{\mu\nu} = (\star R)_{\mu\nu\alpha\beta} K^{\alpha\beta}
\end{equation}
which can be written
\begin{equation}
	\star  {Y}[K]_{\mu\nu} = -(\star R \star)_{\mu\nu\alpha\beta} (\star K)^{\alpha\beta}
\end{equation}
in terms of another CKY 2-tensor $\star K$. Now, when $R_{\mu\nu}=0$, the Riemann tensor and Weyl tensor, $W_{\mu\nu\alpha\beta}$, are equal. Therefore,
\begin{equation}
	\star{Y}[K]_{\mu\nu} = -(\star W \star)_{\mu\nu\alpha\beta} (\star K)^{\alpha\beta} = W_{\mu\nu\alpha\beta} (\star K)^{\alpha\beta} = R_{\mu\nu\alpha\beta} (\star K)^{\alpha\beta} = Y[\star K]_{\mu\nu}
\end{equation}
where we have used the property that $\star W \star = -W$. Hence
\begin{equation}
    \int Y[K] = \int \star Y[\star K]
\end{equation}
As a result, in the absence of sources, Hodge duality doesn't give any new currents: $Y[\star K]$, $\star Y[K]$ and hence $\star Y[\star K]$ all give the same set of currents as the
$Y[K]$.
The equivalence of these charges was checked in Ref.~\cite{Hinterbichler2023GravitySymmetries} for a specific solution of the linearised vacuum field equations.
The dual charges $q[K]$ defined in eq.~\eqref{eq:dual_Penrose_charge} are then related to the Penrose charges in $d=4$ by
\begin{equation}\label{eq:dual_Penrose_charge4}
    q[K] = Q[\star K]
\end{equation}

Similarly, in four dimensions the double dual Riemann tensor $\star R \star$ has many of the same properties as the Riemann tensor in the absence of sources so we could construct the conserved 2-form currents $(\star R \star)_{\mu\nu\alpha\beta} K^{\alpha\beta}$ but, again, this reproduces the same set of currents; this follows from
$(\star R \star)_{\mu\nu\alpha\beta} K^{\alpha\beta}
=(\star R )_{\mu\nu\alpha\beta} (\star K)^{\alpha\beta}
$.

\subsection{Relation of Penrose and secondary currents}

In four dimensions, eq.~\eqref{eq:ADM_ImprovedPenrose_Relation} becomes
\begin{equation}\label{rturtu}
    Y_+[K]_{\mu\nu} = J[k]_{\mu\nu} + \partial^\rho Z[K]_{\mu\nu\rho} 
\end{equation}
where
\begin{equation} \label{ups4}
    Z[K]_{\mu\nu\rho} = 12K_{[\mu\nu} \Gamma\indices{_{\rho\beta]|}^\beta} - \epsilon_{\mu\nu\rho\beta} h\indices{^\beta_\tau}  \tilde{k}^\tau
\end{equation}
with
\begin{equation}\label{eq:ktilde=*dK}
    \tilde{k}_\mu = (\star\tilde{K})_\mu = \frac{1}{3!}\epsilon_{\mu\alpha\beta\gamma} \tilde{K}^{\alpha\beta\gamma}
\end{equation}
Note that in $d=4$, $\tilde{K}$ is a closed CKY 3-form and hence $ \tilde{k}=\star\tilde{K}$ is a Killing vector that is, in general, different from $\hat{K}$, as is explicitly shown in appendix~\ref{app:CKY_Minkowski}. 

As discussed in section~\ref{sec:PenADM}, if $h_{\mu\nu}$ is globally defined then so is $Z[K]$ and the integral of eq.~\eqref{rturtu} gives $Q[K] = Q[k]$ with $k=2(d-3)\hat{K}$, so that the Penrose charges give the ADM charges.
On the other hand, if $h_{\mu\nu}$ is not globally defined then the Penrose charges give the naive ADM charges \eqref{eq:ADM_charge_definition} plus a topological term \eqref{eq:ADM_Electric_Charges}.

Note that comparing with eqs.~\eqref{eq:Z[k]} and \eqref{eq:W[K]} we also have
\begin{equation}
    Y_+[K]_{\mu\nu} = J[k]_{\mu\nu} + \tilde{J}[\tilde{k}]_{\mu\nu} + \partial^\rho W[K]_{\mu\nu\rho} 
\end{equation}
with $W[K]$ given by eq.~\eqref{eq:W[K]}, so that integrating over a 2-cycle $\Sigma_{2}$ gives
\begin{equation}\label{eq:4d_charges}
    Q[K] = Q[k] + \tilde{Q}[\tilde{k}] + \int_{\Sigma_{2}} \dd\star W[K] 
\end{equation}
This suggests that the Penrose charge for a CKY tensor $K$ gives the ADM charge for the Killing vector $k$ given by eq.~\eqref{eq:k=2(d-3)Khat} plus the dual ADM charge for the Killing vector
$\tilde{k}$ given by eq.~\eqref{eq:ktilde=*dK}
plus a topological charge associated with the 3-form $W[K]$.
However, we saw in the last section that for $d>4$ the topological term serves to covariantise the naive ADM charges. Our aim in this section is to analyse the situation for $d=4$.

The improved Penrose 2-form $Y_+[K]$ depends on a CKY tensor $K$, which is given in terms of the constant antisymmetric tensors $\A$, $\B$, $\C$ and $\D$ in eq.~\eqref{eq:CKY_Solution}. 
The results of subsection \ref{trivial_KY_charges} restrict the charges in dimensions $d>4$, so that the only non-trivial Penrose charges arise for the $\B$ and $\D$ tensors. However, in $d=4$ that result does not apply and there are potential Penrose charges for all four tensors $\A$, $\B$, $\C$ and $\D$.

From the results of section~\ref{sec:Penrose2Form}, only the $\B$ and $\D$ terms contribute to the Killing vector $k_\mu = 2(d-3)\hat{K}_\mu$, whereas only the $\C$ and $\D$ terms contribute to the Killing vector $\tilde{k}_\mu = (\star\tilde{K})_\mu$.
Therefore, the Penrose charges for the CKY tensors with only
$\A$ and $\C$ non-zero do not contribute to the charges $Q[k]$ and only appear in the topological terms $Q[\tilde{k}]$ and $\int\dd\star Z[K]$. 

An explicit four-dimensional solution to the linearised Einstein equations is given in Ref.~\cite{Hinterbichler2023GravitySymmetries} and has charges corresponding to all four types of CKY tensor. The parts of the solution which couple to the $\B$- and $\D$-type CKY tensors are globally defined field configurations (they are linearised Schwarzschild and Kerr solutions respectively). However, the parts of the solution which couple to the $\A$- and $\C$-type CKY tensors are, indeed, not globally defined (they are a linearised C-metric and Taub-NUT space respectively). We review part of this solution in section~\ref{sec:examples}.

\subsection{Analysis of Penrose charges in $d=4$}
\label{sec:BreakdownOfPenroseCharges_4d}

We now analyse the Penrose charges for the four  types of CKY tensor in eq.~\eqref{eq:CKY_Solution}.

The $\A$-type CKY tensors are constant 2-forms and so  give $k_\mu=0$ and $\tilde{k}_\mu=0$. Therefore, eq.~\eqref{eq:4d_charges} simplifies to give
\begin{equation}\label{eq:4d_Q[A]}
    Q[\A] = \int_{\Sigma_2} \dd\star Z[\A]
\end{equation}
with
\begin{equation}
    Z[\A]_{\mu\nu\rho} = 12 \A\indices{_{[\mu\nu}} \Gamma\indices{_{\rho\beta]|}^\beta}
\end{equation}
from eq.~\eqref{ups4}. As discussed above,  these charges are non-zero only for non-globally defined $h_{\mu\nu}$ configurations. We will study an example of such a solution in section~\ref{sec:examples}.

The Penrose 2-form for $\A$-type CKY tensors can be written as
\begin{equation}
    Y[\A]_{\mu\nu} = \partial_{[\mu} a_{\nu]}
\end{equation}
where $a$ is the 1-form defined in eq.~\eqref{eq:a_1form}.
The surface $\Sigma_{2}$ is required to be in a region where $R_{\mu\nu}=0$ so that $Q[K]$ is conserved. Therefore, on $\Sigma_{2}$, we have $Y_+[K] = Y[K]$. Then in terms of $a$, we can write the Penrose charge as
\begin{equation}
\label {eleca}
    Q[\A] = \int_{\Sigma_{2}} \star \dd{a}
\end{equation}
which is an electric charge for the 1-form gauge field $a$. 
We  define 2-form charges $\mathbb{M}_{\mu\nu} $ by writing this charge as
\begin{equation}
\label {elecax}
    Q[\A] = \frac 1 2 \A_{\mu\nu} \mathbb{M}^{\mu\nu} 
\end{equation}

The $\B$-type CKY tensors in eq.~\eqref{eq:CKY_Solution} give the constant translation Killing vectors $\frac{1}{2(d-3)} k_\mu=\hat{K}_\mu = -\frac{1}{2}\B_\mu$ and $\tilde{k}_\mu = 0$ and so eq.~\eqref{eq:4d_charges} becomes
\begin{equation}\label{eq:4d_Q[B]}
    Q[\B] = Q[k] + \int_{\Sigma_2} \dd\star Z[\B]
\end{equation}
with
\begin{equation}
    Z[\B]_{\mu\nu\rho} = 12 \B\indices{_{[\mu}} x\indices{_\nu} \Gamma\indices{_{\rho\beta]|}^\beta}
\end{equation}
If $h_{\mu\nu}$ is globally defined, then the integral of the total derivative term $\int\dd\star Z[\B]$ vanishes and  the $\B$-type Penrose charges give the ADM momenta in eq.~\eqref{eq:ADM_charge_definition}. As discussed in section~\ref{sec:PenroseElectricRelation} for $d>4$, if $h_{\mu\nu}$ is not globally defined the total derivative term serves to covariantise the result to give a gauge-invariant definition of the ADM momentum $\mathbb{P}^\mu$:
\begin{equation}
    Q[\B] = -\B_\mu \mathbb{P}^\mu 
\end{equation}

As for  the $\A$-type charges, the Penrose 2-form for the $\B$-type CKY tensors can be written as
\begin{equation}
    Y[\B]_{\mu\nu} = \partial_{[\mu} b_{\nu]} 
\end{equation}
where $b$ is the 1-form defined in eq.~\eqref{eq:b_1form}.
Then the Penrose charge for the $\B$-type CKY tensors is
\begin{equation}
    Q[\B] = \int_{\Sigma_2} \star \dd{b}
\end{equation}
which is  the electric charge for the 1-form $b_\mu$.

The $\C$-type CKY tensors in eq.~\eqref{eq:CKY_Solution} give $k_\mu=0$ while $\tilde{k}_\mu = (\star\C)_\mu$ are constant Killing vectors. Then eq.~\eqref{eq:4d_charges} yields
\begin{equation}\label{eq:4d_Q[C]}
    Q[\C] = \tilde{Q}[\tilde{k}] + \int_{\Sigma_2} \dd\star W[\C]
\end{equation}
with 
\begin{equation}
    W[\C]_{\mu\nu\rho} = 12 x^\sigma \C\indices{_{\sigma[\mu\nu}} \Gamma\indices{_{\rho\beta]|}^\beta}
\end{equation}
The dual momenta $\tilde{Q}[\tilde{k}]$, given in eq.~\eqref{eq:4d_dual_ADM_charges}, are not gauge-invariant in general and the term involving $W[\C]$ serves to covariantise them.
Then the Penrose charge gives a gauge-invariant definition of the dual momenta which we denote $\tilde{\mathbb{P}}^\mu$,
\begin{equation}
    Q[\C] = (\star\C)_\mu \tilde{\mathbb{P}}^\mu 
\end{equation}
In particular, the dual mass (which is related to the four-dimensional NUT charge) is the Penrose charge for $(\star\C)_\mu  = -\delta_\mu ^t$. 

The $\D$-type CKY tensors are the only ones to yield both non-zero $\hat{K}_\mu = \frac{1}{2} k_\mu = \D_{\mu\nu}x^\nu$ and $\tilde{k}_\mu = (\star \D)_{\mu\nu}x^\nu$, so both $Q[k]$ and $\tilde{Q}[\tilde{k}]$ contribute to eq.~\eqref{eq:4d_charges}. For a given $\D$, the two Killing vectors $k_\mu$ and $\tilde{k}_\mu$ are Lorentz Killing vectors giving Lorentz transformations with parameters $\Lambda_{\mu\nu}=\D_{\mu\nu}$ and $\tilde \Lambda_{\mu\nu}=(\star \D)_{\mu\nu}$.
As in section~\ref{sec:PenroseElectricRelation}, the charge $Q[\D]$ gives the ADM angular momentum when $h_{\mu\nu}$ is globally defined and the total derivative term serves to covariantise the definition of the angular momentum when $h_{\mu\nu}$ is not globally defined.
This yields the  improved definition of angular momentum
\begin{equation}
    Q[\D] = \frac 1 2 \D_{\mu\nu} \mathbb{L}^{\mu\nu} 
\end{equation}
that is gauge-invariant even when 
$h_{\mu\nu}$ is not globally defined.

\subsection{Counting and duality}
\label{counting}

In four dimensions, there are 20 CKY tensors, and hence 20 Penrose charges, while there are only 10 Killing vectors, and hence only 10 ADM charges.
This mismatch was one of the puzzles considered in Ref.~\cite{Penrose1982Quasi-localRelativity}.
We have seen that the 4 charges $Q[\B]$ correspond to the 4-momentum and the 6 charges $Q[\D]$ correspond to the angular momentum, and in each of these cases the Killing vectors are proportional to $\hat{K}$.
There is another map from CKY tensors $K$ to Killing vectors, with the Killing vectors given by $\tilde{k} = \star \tilde{K}$, as in eq.~\eqref{eq:ktilde=*dK}, suggesting that the remaining 10 Penrose charges $Q[\A]$, $Q[\C]$ could be related to the Killing vectors $\tilde{k}$.
This turns out to be the case for the $\C$-type CKY tensors but not for the $\A$-type ones.
For the $\C$-type CKY tensors, the vectors $\tilde{k}$ are constant and so are the translation Killing vectors.
However, the $\A$-type CKY tensors have vanishing $k$ and $\tilde{k}$ and so the Penrose charges $Q[\A]$ are not related to any of the charges based on Killing vectors. It is the $\D$-type CKY tensors for which $\tilde k$ are the Lorentz transformation Killing vectors.

In addition to the 10 charges $Q[\B]$, $Q[\D]$ corresponding to the ADM charges there are the 10 charges $Q[\A]$, $Q[\C]$ which we have seen correspond to the 10 KY tensors (that is, the $\A$- and $\C$-type CKY tensors).
As discussed in section \ref{sec:d-3_form_charges}, higher-form symmetries are expected to come in dual pairs with equality between the number of 1-form and $(d-3)$-form symmetries \cite{CasiniCompleteness}.
If $d=4$, the duality is between two 1-form symmetries.
In this case, the four charges $Q[\B]$ are dual to the four charges $Q[\C]$ the six charges $Q[\A]$ are dual to the six charges $Q[\D]$. 
This gives a pairing between the 4-momentum ${\mathbb{P}}^\mu$ and the dual 4-momentum $\tilde{\mathbb{P}}^\mu$ together with a pairing between the angular momentum $ \mathbb{L}_{\mu\nu}$ and the charges $\mathbb{M}_{\mu\nu}$ defined in eq.~\eqref{elecax}.
While we treat the theory classically here, this pairing can also be understood in a canonical quantisation framework \cite{Benedetti2022GeneralizedGraviton, BenedettiNoether}.

It is interesting to compare this with the pairing for $d>4$ in section~\ref{sec:d-3_form_charges} in which $Q[\D]$ is paired with $q[\A]$ and $Q[\B]$ is paired with $q[\B]$. 
In $d=4$, $q[\A]=Q[\star \A]$ is the magnetic charge \eqref{eq:q[A]_mag_charge} for the potential $a$ defined in eq.~\eqref{eq:a_1form}.
Comparing with eq.~\eqref{eleca}, it is also the electric charge for the dual potential $\tilde{a}$ defined by eq.~\eqref{eq:a_1form} with $A$ replaced by $\star A$:
\begin{equation}\label{eq:a_1formt}
   \tilde a_\nu = 2\Gamma\indices{^{\alpha\beta}_{|\nu}} (\star\A)_{\alpha\beta}
\end{equation}
which satisfies
\begin{equation}
     \dd \tilde{a} = \star \dd a
\end{equation}
As $q[\A]=Q[\star \A]$, we can use (\ref{elecax}) to write
\begin{equation}
\label {elecaxq}
    q[\A] = \frac{1}{2} (\star\A)_{\mu\nu} \mathbb{M}^{\mu\nu} 
\end{equation}
so that the pairing of $\mathbb{L}_{\mu\nu}$ with $\mathbb{M}_{\mu\nu}$ indeed corresponds to a pairing of $Q[\D]$ with $q[\A]$.

In $d=4$, from eqs.~\eqref{eq:4d_Q[B]} and \eqref{eq:4d_Q[C]}, 
\begin{equation}
    Q[\B] = -\B_\mu \mathbb{P}^\mu \qc Q[\C] = (\star\C)_\mu \tilde{\mathbb{P}}^\mu
\end{equation}
while
\begin{equation} \label{eq:CKY_SolutionC}
	q[\B] = -\frac{1}{2} \B_\mu \tilde{\mathbb{P}}^\mu \qc q[\C] = -2 (\star \C)_\mu \mathbb{P}^\mu
\end{equation}
so that the pairing  between the 4-momentum
${\mathbb{P}}^\mu$ and the dual 4-momentum $\tilde{\mathbb{P}}^\mu$
can be viewed as a pairing between
$Q[\B]$ and $q[\B]$.

In the absence of sources, the free graviton theory has a dual formulation in terms of a dual graviton $\tilde{h}_{\mu\nu}$ and the dual 4-momentum $\tilde{\mathbb{P}}^\mu$ can be interpreted as the ADM 4-momentum for the dual graviton theory \cite{HullYetAppear}.

\section{Penrose charges in Kaluza-Klein theory}
\label{sec:KaluzaKlein}

\subsection{Linearised gravity on the product of Minkowski space with a torus}

Kaluza-Klein monopole solutions are  BPS states carrying magnetic charges that are an important part of the spectrum of supergravity  and M-theory. In particular, the graviphotons which arise from compactification on tori have magnetic monopole solutions whose uplift to the gravitational theory  carry the gravitational magnetic charges discussed in Refs.~\cite{Hull:1997kt, HullYetAppear}. 

In this section, we analyse the Penrose charges for a background which is 
the product of Minkowski space with a torus and in the next section we will evaluate these charges for linearised Kaluza-Klein monopoles and other solutions.
We will show that the higher-form symmetries of the  graviton and the graviphoton fields in the dimensionally reduced theory are unified from the higher-dimensional perspective, thus giving an interpretation of the uplift of these symmetries.

We focus on solutions of linearised gravity on $\mathcal{M}=\mathbb{R}^{1,D-1}\times T^n$ with $D+n=d$. We denote the coordinates $x^\mu = (x^m,y^i)$ where $y^i$ are periodic coordinates on $T^n$, $y^i\sim y^i+2\pi R_i$, and the $x^m$ are Cartesian coordinates on $\mathbb{R}^{1,D-1}$. The metric on $\mathcal{M}$ is the Minkowski metric $\eta_{\mu\nu}$. 

There are local solutions of the Killing equation \eqref{killlin} of the  form \eqref{eq:Killing_vectors} but those with explicit dependence on $y^i$ are not globally defined on $\mathcal{M}$. 
The globally defined Killing vectors are the constants $k_\mu = V_\mu$ and the vectors $k_\mu $ with $k_m=\Lambda_{mn} x^n$ and $k_i=0$ corresponding to Lorentz transformations on $\mathbb{R}^{1,D-1}$. 
These give rise to conserved charges $P_\mu$ and $L_{mn}$ as before, but now there are no charges $L_{im}$ and $L_{ij}$. The ADM energy for Kaluza-Klein theories was discussed in Ref.~\cite{DeserSoldate1989}.

Similarly, there are local CKY tensors of the form \eqref{eq:CKY_Solution} but only those with no explicit dependence on $y^i$ are globally defined on $\mathcal{M}$.
The CKY tensors on $\mathcal{M}$ are then of the form~\eqref{eq:CKY_Solution} with the only non-zero parameters being $\A_{\mu\nu}$, giving a constant 2-form on $\mathcal{M}$, $\C_{mnp}$, giving a constant 3-form on $\mathbb{R}^{1,D-1}$, and $\B_y$, giving a constant scalar on $\mathbb{R}^{1,D-1}$. From these CKY tensors we can construct corresponding Penrose currents and charges, as before. 
Note that, as there are no CKY tensors with $\B_m\neq0$ or $\D_{mn}\neq0$, it appears that $P_m$ and $L_{mn}$ cannot be expressed as Penrose charges. We will discuss below how these charges do in fact have a covariant expression in the dimensionally reduced theory.

Of particular interest are configurations in which the graviton field is independent of the toroidal coordinates, 
\begin{equation}
\label{dhiso}
\partial_i h_{\mu\nu}=0
\end{equation}
For these, as we shall see, more conserved charges can be defined. Such configurations can be dimensionally reduced to a field theory in $D$ dimensions and the charges are most easily analysed in the dimensionally reduced theory. We analyse the currents and charges in the dimensionally reduced theory in the following subsections, and then examine their lift back up to $d$ dimensions.

\subsection{Kaluza-Klein Ansatz}

We take a Kaluza-Klein Ansatz of the following form for the $d$-dimensional graviton $h_{\mu\nu}$
\begin{equation}\label{eq:KK_ansatz}
    h_{\mu\nu} \quad \longrightarrow \quad h_{mn} = \bar{h}_{mn} - \frac{2}{D-2} \eta_{mn}\phi \qc h_{mi} = 2A_m^{(i)} \qc h_{ij} = 2\phi^{(ij)}
\end{equation}
where $\phi \equiv \sum_{i=1}^n \phi^{(ii)}$.
We see that the $D$-dimensional theory is governed by a $D$-dimensional graviton $\bar{h}_{mn}$, $n$ one-form graviphoton fields $A^{(i)}_m$, and $n(n+1)/2$ scalars $\phi^{(ij)}$, which are arranged in a symmetric $n\times n$ matrix. 
We take all fields to be independent of the compact dimensions; that is, $\partial_i h_{\mu\nu}=0$. It immediately follows that the $d$-dimensional linearised Riemann tensor satisfies
\begin{equation}\label{eq:dy_R=0}
    \partial_i R_{\mu\nu\rho\sigma} = 0 
\end{equation}
The components of the linearised Riemann tensor are
\begin{equation}\label{eq:Riemann_components_S1}
\begin{split}
    R_{mnpq} &= \bar{R}_{mnpq} - \frac{2}{D-2} \left( \eta_{m[p}\partial_{q]}\partial_n \phi - \eta_{n[p} \partial_{q]}\partial_m \phi \right) \\ 
    R_{mnpi} &= \partial_p F^{(i)}_{mn} \\ 
    R_{minj} &= \partial_m \partial_n \phi^{(ij)}
\end{split}
\end{equation}
where $\bar{R}_{mnpq} = -2 \partial_{[m} \bar{h}_{n][p,q]}$ is the curvature of the $D$-dimensional graviton and $F^{(i)}_{mn} = \partial_m A^{(i)}_n - \partial_n A^{(i)}_m$ is the field strength of the $i^{\text{th}}$ graviphoton.

The Ansatz in eq.~\eqref{eq:KK_ansatz} is chosen so that $G_{mn} = \bar{G}_{mn}$, where $\bar{G}_{mn}$ is the Einstein tensor for the $D$-dimensional graviton $\bar{h}_{mn}$. Then the $d$-dimensional field equations $G_{\mu\nu}=T_{\mu\nu}$ imply the $D$-dimensional equations
\begin{equation}
    \bar{G}_{mn} = T_{mn} \qc \partial^p F_{pm}^{(i)} = T_{mi}\qc \partial^m \partial_m \phi^{(ij)} = T_{ij} - \frac{1}{D+n-2}\eta_{ij} T\indices{^\mu_\mu}
\end{equation}
Denoting the various components of the energy-momentum tensor as
\begin{equation}
    T_{mn} \qc T_{mi} \equiv j^{(i)}_m \qc T_{ij} \equiv f^{(ij)} + \frac{1}{D+n-2}\eta_{ij} T\indices{^\mu_\mu}
\end{equation}
the equations of motion for the $D$-dimensional fields can be written
\begin{equation}\label{eq:EoM_S1}
\begin{split}
    \bar{G}_{mn} &= T_{mn}\\
    \partial^p F^{(i)}_{pm} &= j^{(i)}_m \\
    \partial_m \partial^m \phi^{(ij)} &= f^{(ij)}
\end{split}
\end{equation}
Therefore, in the $D$-dimensional theory we see that $T_{mn}$  is the source for the graviton $\bar{h}_{mn}$, the $T_{mi}=j^{(i)}_m$ are electric sources for the graviphotons, and the scalars are sourced by both $T_{ij}$ components and the trace $T\indices{^\mu_\mu}$.

\subsection{2-form currents in the absence of sources}
\label{sec:KaluzaKlein_currents}

As in section~\ref{sec:Penrose2Form}, we will first consider the conserved 2-form currents in a region where $\bar{R}_{mn}=0$ and $\partial^p F^{(i)}_{pm}=0$. If the field equations \eqref{eq:EoM_S1} hold, this is a region where $T_{\mu\nu}=0$.
We can build two such 2-forms in the $D$-dimensional theory,
\begin{equation}\label{eq:Y[K]_and_Z[xi]}
    Y[\bar{K}]_{mn} = \bar{R}_{mnpq} \bar{K}^{pq}\qc X[\xi]_{mn} = R_{mnpi} \xi^{(i)p}
\end{equation}
for a 2-form $\bar{K}_{mn}$ and $n$ one-forms $\xi^{(i)}_p$ in $D$ dimensions. 
(2-tensors of the form $R_{minj}\vartheta^{(ij)} = \partial_m \partial_n \phi^{(ij)} \vartheta^{(ij)}$ are symmetric in the $m,n$ indices and do not lead to charges of the type discussed here.)
By the same arguments as in section~\ref{sec:Penrose2Form}, 
the 2-form current $Y[\bar{K}]$ is conserved in regions where $\bar{R}_{mn}=0$ provided that $\bar{K}_{mn}$ is a CKY tensor on $\mathbb{R}^{1,D-1}$. The conservation of $X[\xi]$ follows similar lines. We have
\begin{equation}\label{eq:Z_conservation}
    \partial^n X[\xi]_{mn} = \partial^n R_{mnpi} \xi^{(i)p} + R_{mnpi} \partial^n \xi^{(i)p}
\end{equation}
The first term vanishes when $\partial^p F^{(i)}_{pm}=0$, viz.
\begin{equation}
    \partial^n R_{mnpi} = \partial^n \partial_p F^{(i)}_{mn} = 0
\end{equation}
The other term in eq.~\eqref{eq:Z_conservation} similarly vanishes provided that $\xi^{(i)}$ satisfies
\begin{equation}
    \partial^n \xi^{(i)p} = \eta^{np} \hat{\xi}^{(i)}
\end{equation}
where
\begin{equation}
    \hat{\xi}^{(i)} = \frac{1}{D} \partial_p \xi^{(i)p}
\end{equation}
This is the condition for the $\xi^{(i)}$ to be closed conformal Killing vectors (closed CKVs) on $D$-dimensional Minkowski space. The general solution for $\xi^{(i)}$ on $\mathbb{R}^{1,D-1}$ is
\begin{equation}\label{eq:xi_solution}
    \xi^{(i)}_m = n^{(i)}_m + m^{(i)} x_m
\end{equation}
where the $n^{(i)}$ are constant one-forms and $m^{(i)}$ are constants. Then
\begin{equation}\label{eq:xihat}
    \hat{\xi}^{(i)} = m^{(i)}
\end{equation}
is a constant.

\subsection{Improvement in the presence of sources}

We now relinquish the constraint that $\bar{R}_{mn}=0$ and $\partial^p F^{(i)}_{pm}=0$.
As seen in section~\ref{sec:PenroseImprovement}, in the presence of sources we can add improvement terms to the 2-form currents which involve Ricci tensors. These improvements were useful in writing the relations between the Penrose charges and the ADM charges in section~\ref{sec:PenroseElectricRelation}.
We now give the relevant improvement terms for Kaluza-Klein solutions, the conserved 2-form currents for which are given in eq.~\eqref{eq:Y[K]_and_Z[xi]} with $\bar{K}_{mn}$ a CKY tensor and $\xi^{(i)}_m$ closed CKVs. 

The 2-form current $Y[\bar{K}]_{mn}$ is the $D$-dimensional Penrose 2-form current for the CKY tensor $\bar{K}_{pq}$ on $\mathbb{R}^{1,D-1}$.
Therefore, the improvement terms are the $D$-dimensional version of those considered previously in eq.~\eqref{eq:Def_ImprovedPenrose}, 
\begin{equation}
    Y_+[\bar{K}]_{mn} = \bar{R}_{mnpq} \bar{K}^{pq} + 4\bar{R}\indices{^p_{[m}} \bar{K}_{n]p} + \bar{R} \bar{K}_{mn} 
\end{equation}
However, $X[\xi]_{mn}$ is of a different form. We find that the relevant improved 2-form is 
\begin{align}
    X_+[\xi]_{mn} = R_{mnpi} \xi^{(i)p} + 2R\indices{^p_{ip[m}} \xi^{(i)}_{n]}
\end{align}
Their divergences are given by 
\begin{align}
    \partial^n Y_+[\bar{K}]_{mn} &= 2(D-3) \bar{G}_{mn} \hat{\bar{K}}^n = j[\bar{k}]_m \label{eq:div_Y+}\\
    \partial^n X_+[\xi]_{mn} &= \ell^{(i)} \partial^n F^{(i)}_{mn} \label{eq:div_Z+}
\end{align}
where $j[\bar{k}]_m = \bar{G}_{mn} \bar{k}^n$, with $\bar{G}_{mn}$ the Einstein tensor of the $D$-dimensional graviton, and we have set 
\begin{equation}\label{eq:li_def}
    \frac{1}{2(D-3)} \bar{k}_m = \hat{\bar{K}}_m \equiv \frac{1}{D-1} \partial^n \bar{K}_{nm} \qc \ell^{(i)} = -(D-3) \hat{\xi}^{(i)} 
\end{equation}
The right-hand side of both equations~\eqref{eq:div_Y+} and \eqref{eq:div_Z+} vanishes on-shell in regions where $T_{\mu\nu}=0$ from the field equations~\eqref{eq:EoM_S1}, as was required by the arguments of the previous subsection.

As seen in previous sections, this gives a relation between the 2-form currents $Y_+[\bar{K}]$ and $X_+[\xi]$ and the 2-forms $F^{(i)}_{mn}$ and $J[\bar{k}]_{mn}$ (whose divergence gives the 1-form current $j[\bar{k}]_m$).
The relation is found in the same manner as before. 
We have a 2-form secondary current $J[\bar{k}]_{mn}$, given by eq.~\eqref{eq:J[k]_def} for the $D$-dimensional graviton instead of the $d$-dimensional one, such that
\begin{equation}
    j[\bar{k}]_m = \partial^n J[\bar{k}]_{mn}
\end{equation}
Then by the Poincar\'{e} lemma, eqs.~\eqref{eq:div_Y+} and \eqref{eq:div_Z+} imply relations between the 2-forms $Y_+[\bar{K}]$ and $X_+[\xi]$, and the 2-forms $J[\bar{k}]$ and $F^{(i)}$. Indeed, we finds 
\begin{align}
    Y_+[\bar{K}]_{mn} &= J[\bar{k}]_{mn} + \partial^p \bar{Z}_{mnp} \label{eq:Y+=J_S1} \\
    X_+[\xi]_{mn} &= \ell^{(i)} F^{(i)}_{mn} + \partial^p \Delta_{mnp} \label{eq:Z+=F_S1} 
\end{align}
where $\bar{Z}$ and $\Delta$ are 3-forms given by
\begin{align}
    \bar{Z}_{mnp} &= 12 \bar{K}_{[mn} \bar{\Gamma}\indices{_{pq]|}^q} + 4 \bar{h}\indices{^q_{[m}} \tilde{\bar{K}}_{npq]} \label{eq:Upsilon_D} \\
    \Delta_{mnp} &= 3F^{(i)}_{[mn}\xi^{(i)}_{p]} \label{eq:Delta}
\end{align}
The form of $\bar{Z}$ is the same as that of $Z$ in eq.~\eqref{eq:UpsilonGeneral}, here evaluated for the $D$-dimensional graviton.

Notice that when $\star X_+[\xi]$ is integrated over a cycle, the total derivative term $\partial^p \Delta_{mnp}$ will vanish as $\Delta$ depends on $A^{(i)}_m$ only through the curvature $F^{(i)}_{mn}$, which is globally defined.

\subsection{Analysis of the charges}

For the $D$-dimensional dimensionally reduced theory, conserved charges are constructed by integrating $\star Y_+[\bar{K}]$ and $\star X_+[\xi]$ over a $(D-2)$-dimensional surface $\Sigma_{D-2}$. We now require that $\Sigma_{D-2}$ is contained in a region of $\mathbb{R}^{1,D-1}$ where $\bar{R}_{mn}=0$ and $\partial^p F_{pm}^{(i)}=0$, so that the following charges are conserved in the sense that $\Sigma_{D-2}$ can be arbitrarily deformed within that region.
We define
\begin{align}
    Q[\bar{K}] &= \int_{\Sigma_{D-2}} \bar{\star} Y_+[\bar{K}] \label{eq:Q[Kbar]}\\
    Q[\xi] &= \int_{\Sigma_{D-2}} \bar{\star} X_+[\xi] \label{eq:Q[xi]}
\end{align}
where $\bar{\star}$ is the Hodge dual on $\mathbb{R}^{1,D-1}$.

These charges can be written in terms of $J[\bar{k}]$ and $F^{(i)}$ using eq.~\eqref{eq:Z+=F_S1}.
Recalling that the $\Delta$ contribution in eq.~\eqref{eq:Z+=F_S1} vanishes when integrated, we have
\begin{align}
    Q[\bar{K}] &= Q[\bar{k}] + \int_{\Sigma_{D-2}} \dd\,\bar{\star} \bar{Z} \label{eq:Q[K]_S1_Q[k]} \\ 
    Q[\xi] &= \ell^{(i)} \int_{\Sigma_{D-2}} \bar{\star} F^{(i)} \label{eq:Q[xi]_S1}
\end{align}
where 
\begin{align}
    Q[\bar{k}] &\equiv \int_{\Sigma_{D-2}} \bar{\star} J[\bar{k}]
\end{align}
are the ADM charges of the $D$-dimensional graviton.
Eq.~\eqref{eq:Q[K]_S1_Q[k]} is the $D$-dimensional version of the result we had previously for $d$ dimensions, and the analysis of the previous sections then immediately applies here, with different results for the cases $D=4$ and $D>4$.

The new feature of these results is the set of $Q[\xi]$ charges. Eq.~\eqref{eq:Q[xi]_S1} equates these charges to the electric charges for the graviphotons. 
These generate electric $U(1)$ 1-form symmetries \cite{Gaiotto2015GeneralizedSymmetries}. Recalling eqs.~\eqref{eq:xihat} and \eqref{eq:li_def}, we note that this charge is associated with the $m^{(i)}$-type terms in $\xi^{(i)}$. The $n^{(i)}$-type terms in eq.~\eqref{eq:xi_solution} contribute to $X_+[\xi]$ only via $\Delta$ in eq.~\eqref{eq:Delta}, which vanishes when integrated.

As a result, the charges associated with the $D$-dimensional fields in the Kaluza-Klein reduction of $d$-dimensional linearised gravity on an $n$-torus include the ADM charges of the $D$-dimensional graviton, the electric charges of the graviphotons, together with the magnetic charges for the $D$-dimensional graviton arising when $\bar{h}_{mn}$ is a non-globally defined gauge field configuration.

\subsection{Dual charges}
\label{sec:d-3_form_charges_S1}

As noted in section~\ref{sec:d-3_form_charges}, in $d$ dimensions there are dual charges in the linearised graviton theory which are supported on 2-dimensional surfaces and which are parameterised by closed CKY tensors $\sigma$. We now show that the same is true of the Kaluza-Klein theory.

For the case  $D=4$, we find that $Y[\bar{K}]$ is closed and co-closed for all CKY tensors $\bar{K}_{mn}$ and $X[\xi]$ is closed and co-closed for all closed CKVs $\xi^{(i)}_m$. This results from the duality properties of CKY tensors in four dimensions discussed in section~\ref{sec:four_dimensions}. As found previously, the dual charges $q[K]$ are already included in the set of Penrose charges $Q[K]$.

For $D>4$, we find that $Y[\bar{K}]_{mn}$ is closed when $\bar{K}_{mn}=\bar{\sigma}_{mn}$ is a closed CKY tensor. We also finds that $X[\xi]$ is closed for all closed CKVs $\xi^{(i)}$. Therefore, we can integrate over a 2-cycle $\Sigma_2$ in a region of $\mathbb{R}^{1,D-1}$ away from sources to give conserved charges. 
Integrating $Y[\bar{\sigma}]$ over a 2-cycle  gives the charge $q[\bar{\sigma}]$ for the $D$-dimensional graviton which is familiar from section~\ref{sec:d-3_form_charges}. 
Integrating $X[\xi]$ over a 2-cycle yields new charges. We write
\begin{equation}\label{eq:R_xi_rewriting}
    R_{mnpi}\xi^{(i)p} = \partial_p F^{(i)}_{mn} \xi^{(i)p} = -\frac{1}{2} \hat{\xi}^{(i)} F^{(i)}_{mn} - \frac{1}{2} \partial\indices{_{[m}} \left( F^{(i)}_{n]p} \xi^{(i)p} \right)
\end{equation}
where we have used eq.~\eqref{eq:Riemann_components_S1} in the first equality and $\partial_{[p}F^{(i)}_{mn]}=0$ in the second. Integrating over a 2-cycle, the final term vanishes by Stokes' theorem as it is the exterior derivative of a globally defined 1-form. Therefore, integrating $Y[\bar{\sigma}]$ and $X[\xi]$ over a 2-cycle yields charges
\begin{equation}\label{eq:d-3_charges_S1}
\begin{split}
    q[\bar{\sigma}] &= \int_{\Sigma_2} Y[\bar{\sigma}]\\
    q[\xi] &= \int_{\Sigma_2} X[\xi] =  - \frac{1}{2} \hat{\xi}^{(i)} \int_{\Sigma_2} F^{(i)}
\end{split}
\end{equation}
The charges $\int F^{(i)}$ are the magnetic charges for the graviphotons, which generate $(D-3)$-form symmetries \cite{Gaiotto2015GeneralizedSymmetries}.
Recall from eq.~\eqref{eq:xihat} that $\hat{\xi}^{(i)}$ is only non-zero for $m^{(i)}$-type closed CKVs $\xi^{(i)}_m$, so the constant $n^{(i)}$-type terms in eq.~\eqref{eq:xi_solution} yield vanishing charges as they contribute only to the total derivative term in eq.~\eqref{eq:R_xi_rewriting}, which integrates to zero.

Finally, we return to the expectation that higher-form symmetries should come in dual pairs that was discussed in section~\ref{sec:d-3_form_charges}. We have seen that the Kaluza-Klein type solutions considered in this section have charges supported on codimension-2 cycles and dual charges supported on 2-dimensional cycles (generating 1-form and $(D-3)$-form symmetries respectively).
When $D=4$, both types of cycles are 2-dimensional and the two types of charges are not independent. They are generated by charges corresponding to four-dimensional CKY tensors $K_{mn}$ and four-dimensional closed CKVs $\xi_m^{(i)}$. The constant closed CKVs $\xi^{(i)}_m = n^{(i)}_m$ produce vanishing charges. Therefore, there are $20+n$ non-trivial charges in $D=4$.

When $D>4$, the charges $q[\bar{\sigma}]$ defined on 2-cycles are built only from the closed $D$-dimensional CKY tensors and the charges $q[\xi]$ are built only from the $m^{(i)}$-type closed CKVs. Again, the charges associated with the constant $n^{(i)}$-type closed CKVs are trivial. The Penrose charges $Q[K]$ are built from the $\B$- and $\D$-type $D$-dimensional CKY tensors and the charges $Q[\xi]$ are built from the $m^{(i)}$-type closed CKVs. The remaining CKY tensors and the constant closed CKVs produce trivial charges. So, again, there is the expected duality between the dimension-2 and codimension-2 charges (generating $(D-3)$-form and 1-form symmetries respectively), with $\frac{1}{2}D(D+1)+n$ of each. 

\subsection{Uplift to $d$ dimensions}
\label{sec:KK_uplift}

We now ask whether these 2-form currents in $D$ dimensions and their associated charges have a unified origin in $d$ dimensions. 
First we introduce a 2-form $\mathscr{K}$ on  $\mathcal{M}$ with components
\begin{equation}\label{eq:Kscr_K_xi}
    \mathscr{K}_{mn} = \bar{K}_{mn} \qc \mathscr{K}_{mi} = \xi^{(i)}_m/2
\end{equation}
with $\mathscr{K}_{ij}$ unrestricted.
($\mathscr{K}_{ij}$ does not enter in the following analysis and can be chosen arbitrarily, e.g.\ $\mathscr{K}_{ij}=0$.)
We will consider the object
\begin{equation}\label{eq:Y[Kscr]=Y[K]+Z[xi]+phi}
    Y[\mathscr{K}]_{mn} = Y[\bar{K}]_{mn} + X[\xi]_{mn} 
\end{equation}
which are the $mn$ components of the $d$-dimensional 2-form
\begin{equation}\label{eq:Y[Kscr]_def}
    Y[\mathscr{K}]_{\mu\nu} = (R_{\mu\nu\rho\sigma} + V_{\mu\nu\rho\sigma}) \mathscr{K}^{\rho\sigma}
\end{equation}
where the only non-zero components of $V_{\mu\nu\rho\sigma}$ are
\begin{equation}
    V_{mnpq} = \frac{2}{D-2} \left( \eta_{m[p}\partial_{q]}\partial_n \phi - \eta_{n[p} \partial_{q]}\partial_m \phi \right)
\end{equation}
This is seen as follows. We note that the first of eqs.~\eqref{eq:Riemann_components_S1} can be written
\begin{equation}
    R_{mnpq} = \bar{R}_{mnpq} - V_{mnpq}
\end{equation}
from which it follows that
\begin{equation}
    Y[\mathscr{K}]_{mn} = \bar{R}_{mnpq} \mathscr{K}^{pq} + 2 R_{mnpi} \mathscr{K}^{pi}
\end{equation}
Upon comparison with eq.~\eqref{eq:Y[K]_and_Z[xi]}, this gives eq.~\eqref{eq:Y[Kscr]=Y[K]+Z[xi]+phi}.

The charge corresponding to a given $\mathscr{K}$ is found by integrating $\star Y[\mathscr{K}]$ over a codimension-2 cycle in a region of $\mathcal{M}$ away from sources. We take this cycle to fully wrap the $n$-torus
\begin{equation}
    \Sigma_{d-2} = \Sigma_{D-2} \times T^n
\end{equation}
so that the charges built from $Y[\mathscr{K}]$ can be related to those of $Y[\bar{K}]$ and $X[\xi]$ in $\mathbb{R}^{1,D-1}$. For example, we may take $\Sigma_{D-2}$ to be the $(D-2)$-sphere at spatial infinity in $\mathbb{R}^{1,D-1}$.
We then have
\begin{equation}\label{eq:Q[Kscr]=Q[k]+Q[xi]}
    Q[\mathscr{K}] \equiv \int_{\Sigma_{d-2}} \star Y[\mathscr{K}] = \text{vol}_{T^n} \left( Q[\bar{K}] + Q[\xi] \right)
\end{equation}
where $Q[\bar{K}]$ and $Q[\xi]$ were defined in eqs.~\eqref{eq:Q[Kbar]} and \eqref{eq:Q[xi]} respectively, and
\begin{equation}
    \text{vol}_{T^n} = \prod_{i=1}^n 2\pi R_i
\end{equation}
is the volume of the torus. Therefore we see that the charges generating the higher-form symmetries in the $D$-dimensional theory have an uplift to the $d$-dimensional theory. 

Then, from eqs.~\eqref{eq:Q[K]_S1_Q[k]} and \eqref{eq:Q[xi]_S1}, the charges on the right-hand side of eq.~\eqref{eq:Q[Kscr]=Q[k]+Q[xi]} are related to the ADM and dual ADM charges of the $D$-dimensional graviton so eq.~\eqref{eq:Q[Kscr]=Q[k]+Q[xi]} gives a covariant $d$-dimensional origin for these charges.

Note that the terms involving $\phi$ in eq.~\eqref{eq:Y[Kscr]_def} can be absorbed into a field redefinition of the $d$-dimensional graviton.
We define $h'_{\mu\nu}$ to have components 
\begin{equation}
    h'_{mn} = h_{mn} + \frac{2}{D-2} \eta_{mn} \phi = \bar{h}_{mn} \qc h'_{mi} = h_{mi} \qc h'_{ij} = h_{ij}
\end{equation}
where $\bar{h}_{mn}$ is defined in eq.~\eqref{eq:KK_ansatz}. The curvature tensor $R'$ for $h'$ is
\begin{equation}
    R'_{\mu\nu\rho\sigma} =R_{\mu\nu\rho\sigma} +V_{\mu\nu\rho\sigma}
\end{equation}
so that now $Y[\mathscr{K}]$ in eq.~\eqref{eq:Y[Kscr]_def} can be written
\begin{equation}\label{eq:Y[Kscr]_R'}
    Y[\mathscr{K}]_{\mu\nu} = R'_{\mu\nu\rho\sigma} \mathscr{K}^{\rho\sigma}
\end{equation}
which is of the form of a Penrose current.

However, the current is not quite a Penrose current of the type discussed in previous sections as $\mathscr{K}$ is not a CKY tensor in $d$ dimensions.
Instead, it satisfies
\begin{equation}\label{eq:CKY_equationd}
    \partial_p \mathscr{K}_{mn} =\tilde  { \mathscr{K}}_{pmn} + 2 \eta_{p[m} \hat {\mathscr{K}}_{n]}
    \qc \partial_i \mathscr{K}_{mn} =0
\end{equation}
where 
\begin{equation} \label{eq:Khat_Ktilde_defd}
   \hat{\mathscr{K}}_m \equiv \frac{1}{D-1}\partial^n \mathscr{K}_{nm} \qc \tilde{\mathscr{K}}_{mnp} \equiv \partial_{[m} \mathscr{K}_{np]}
\end{equation}
and
\begin{equation}
    \partial_n \mathscr{K}_{pi} = \eta_{np} \hat{\mathscr{K}}_{i} \qc  \partial_j \mathscr{K}_{pi} = 0
\end{equation}
where
\begin{equation}\label{eq:Kscr_hat_i_def}
   \hat{\mathscr{K}}_{i} = \frac{1}{D} \partial^p\mathscr{K}_{pi}
\end{equation}
That is, its $\mathscr{K}_{mn}$ components are those of a $D$-dimensional CKY tensor, while its $\mathscr{K}_{mi}$ components are those of $n$ closed CKVs in $D$-dimensions.
This leads to the conservation of the current \eqref{eq:Y[Kscr]_R'} as
$R'_{\mu\nu\rho\sigma}$ satisfies
\begin{equation}
\eta ^{pq} R'_{\mu p\nu q}=0
\end{equation}
in the absence of sources, from eq.~\eqref{eq:EoM_S1}.
The fact that its trace with the $D$-dimensional metric $\eta_{pq}$ vanishes allows the possibility of terms involving $\eta_{pq}$ on the right hand side of $\partial_n \mathscr{K}_{\mu\nu}$ in the conditions for conservation of the current.

The analysis of section \ref{sec:Penrose2Form} can be extended to allow for such possibilities in which restricting to a special set of configurations allows a more general set of conserved charges.
For configurations in which the curvature is traceless with respect to some symmetric tensor $\Pi^{\mu\nu}$, i.e.
\begin{equation}\label{eq:pi_R}
    \Pi^{\mu\nu} R_{\mu\rho\nu\sigma} = 0
\end{equation}
the condition for the current \eqref{eq:Penrose_2form} to be conserved when $R_{\mu\nu}=0$ is that
\begin{equation}\label{eq:CKY_equationp}
    \partial_\rho K_{\mu\nu} = \tilde{K}_{\rho\mu\nu} + 2 \eta_{\rho[\mu} \hat K_{\nu]} + 2 \Pi_{\rho[\mu}  K'_{\nu]}
\end{equation}
for some 1-forms $\hat{K}_\mu$ and $K'_\mu$.
Here $\Pi_{\mu\nu} = \eta_{\mu\rho} \eta_{\nu\sigma} \Pi^{\rho\sigma}$ 
and $\tilde{K}_{\rho\mu\nu} \equiv \partial_{[\rho} K_{\mu\nu]}$, as before.
The one-forms $\hat{K}_\mu$ and $K'_\mu$ are then constrained by taking traces with $\eta^{\mu\nu}$, $\Pi^{\mu\nu}$ and
$(\Pi^N)^{\mu\nu}$ (where, e.g., for $N=2$, $(\Pi^2)^{\mu\nu} = \Pi^{\mu\rho}\eta_{\rho\sigma}\Pi^{\sigma\nu}$).

For the case analysed above, $R'_{\mu\nu\rho\sigma}$ satisfies
\begin{equation}
    \eta^{pq} R'_{\mu p\nu q} = 0
\end{equation}
on-shell, which can be written as in eq.~\eqref{eq:pi_R} with
\begin{equation}
    \Pi _{\mu\nu} = \begin{pmatrix}
    \eta _{mn} & 0 \\
    0 & 0
    \end{pmatrix}
\end{equation}
However its Ricci tensor doesn't vanish in general: $R'_{\mu\nu}=\eta ^{\rho\sigma} R'_{\mu\rho\nu\sigma}\ne 0$.
Then the condition for the conservation of the 2-form \eqref{eq:Penrose_2form} is
\begin{equation}\label{eq:CKY_equationpp}
    \partial_\rho K_{\mu\nu} = \tilde{K}_{\rho\mu\nu} + 2 \Pi_{\rho[\mu} K'_{\nu]}
\end{equation}
This then implies that
\begin{equation}
    \Pi^{\rho\mu} \partial_\rho K_{\mu\nu} = DK'_{\nu} - \Pi_{\nu\rho} K'^{\rho}
\end{equation}
giving
\begin{equation} \label{asfgf}
     {K}'_m \equiv \frac{1}{D-1}\partial^n K_{n m}
\end{equation}
and
\begin{equation}
   {K}'_{i} = \frac{1}{D} \partial^p K_{pi}
\end{equation}
in agreement with eqs.~\eqref{eq:Khat_Ktilde_defd} and \eqref{eq:Kscr_hat_i_def} on replacing $\mathscr{K}_{\mu\nu}\to K_{\mu\nu}$ and $\hat{\mathscr{K}}_{\mu} \to K'_{\mu}$.

\section{Penrose charges for various solutions}
\label{sec:examples}

In this section we evaluate the charges discussed in the previous sections for various  solutions of the linearised Einstein equations. 
The linearisation arises from writing the full metric as $g_{\mu\nu}=\bar g_{\mu\nu}+h_{\mu\nu}$
with $\bar g_{\mu\nu}$ a solution of the Einstein equations and then the Fierz-Pauli equations for $h_{\mu\nu}$ are the terms in the Einstein equation for $g_{\mu\nu}$ that are linear in $h_{\mu\nu}$, and involve the metric connection $\bar \nabla_\mu$ for the background metric $\bar g_{\mu\nu}$. In this paper, we take $\bar g_{\mu\nu}$ to be the Minkowski metric. We have so far used  Cartesian coordinates for which $\bar \nabla_\mu=\partial _\mu$, but in this section we also use spherical polar coordinates.

In the linearised theory, linear superpositions of solutions are again solutions, while of course this is not the case in the non-linear Einstein theory. 

We discuss a solution with electric gravitational charges (i.e. ADM mass and angular momentum) which correspond to the $\B$- and $\D$-type CKY tensors, a solution with magnetic gravitational charge corresponding to the $\C$-type CKY tensors, a solution which carries the $(d-3)$-form symmetry charge $q[K]$, and finally a solution which carries the $\A$-type Penrose charge.

\subsection{Five-dimensional linearised Myers-Perry black hole solution}

An example of a solution which carries electric gravitational charges is the linearised five-dimensional Myers-Perry black hole metric. This solution is not well-defined at the origin of the spatial $\mathbb{R}^4$, so we are considering its linearisation around a background space which is $\mathbb{R}^{1,4}$ with the line $r=0$ removed. We denote this space $(\mathbb{R}^{1,4})^{\bullet}$ and parameterise it by spherical coordinates $\{t,r,\theta,\phi,\psi\}$.

The linearised Myers-Perry metric has non-vanishing components
\begin{align}\label{eq:MyersPerry_h}
    h_{tt} = h_{rr} = \frac{M}{r^2}\qc h_{t\phi} = \frac{J_1 \sin^2 \theta}{r^2}\qc h_{t\psi} = \frac{J_2 \cos^2 \theta}{r^2}
\end{align}
where $M$ is the mass parameter of the black hole and $J_1,J_2$ are parameters corresponding to the rotation of the black hole in the $\phi$ and $\psi$ directions respectively (that is, in the planes spanned by the Cartesian coordinates $x^1,x^2$ and $x^3,x^4$ respectively).

Calculating the Penrose charges for the different types of CKY tensors yields
\begin{equation}
    Q[\A] = 0\qc Q[\B] = -12 \pi^2 M \B_0\qc Q[\C] = 0\qc Q[\D] = -16 \pi^2 (J_1 \D_{12} + J_2 \D_{34})
\end{equation}
where $\B_0$, $\D_{12} $ and $ \D_{34}$ are the components in Cartesian coordinates.
Firstly, we note that the charges $Q[\A]$ and $Q[\C]$ vanish. This is in agreement with the discussion of section~\ref{trivial_KY_charges} as the $\A$- and $\C$-type CKY tensors are KY tensors and so do not contribute to the Penrose charges when $d>4$.

The Penrose charge $Q[\B _0]$ is the ADM mass, in agreement with eq.~\eqref{eq:B_and_D_type_ADM_link} as the $\B$-type CKY tensors correspond to translational Killing vectors. The Penrose charge $Q[\D]$ gives the two independent ADM angular momenta, which also agrees with eq.~\eqref{eq:B_and_D_type_ADM_link} as the $\D$-type CKY tensors correspond to rotational Killing vectors. Note that the rotation parameters $J_1$ and $J_2$ are picked out by the components of the CKY parameter $\D_{\mu\nu}$ in the planes orthogonal to the rotation axis. 

For the solution in eq.~\eqref{eq:MyersPerry_h}, the total derivative contributions to the Penrose charges vanish so the Penrose charges give precisely the ADM charges in this case.

\subsection{Linearised Lorentzian Taub-NUT}
\label{sec:LorTaubNut}

We now consider an example of a graviton configuration in four dimensions with non-trivial topology that carries 
magnetic gravitational charge.
We denote the coordinates on $\mathbb{R}^{1,3}$ by $x^m = (t,x^\alpha)$ with $\alpha=1,2,3$.
We will consider a background spacetime $(\mathbb{R}^{1,3})^{\bullet}$ which, similarly to the previous section, is defined as $\mathbb{R}^{1,3}$ with the line given by $x^\alpha=0$ removed.
Consider the Ansatz
\begin{equation}
    h_{\alpha t} = 2A_\alpha(x^\beta)
\end{equation}
where $A_\alpha$ is a 1-form connection on $\mathbb{R}^3\setminus\{0\}$ which is independent of $t$ with field strength
\begin{equation}\label{eq:F=dA}
    F_{\alpha\beta} = \partial_\alpha A_\beta - \partial_\beta A_\alpha
\end{equation}
We choose $A$ so that the field strength is given by a potential $V$ with 
\begin{equation}\label{eq:LorTaubNUT_F}
    F_{\alpha\beta} = \epsilon_{\alpha\beta\gamma} \partial_\gamma V
\end{equation}
and we choose $V$ to be  
\begin{equation}\label{eq:LorTaubNUT_X}
    V(x^\alpha) = - \frac{N}{\abs{x}}
\end{equation}
corresponding to a source at $x^\alpha =0$ with strength $N$,  referred to as the NUT charge.
The non-zero components of the curvature tensor are then
\begin{equation}\label{eq:LorTaubNUT_Riemann}
    R_{\alpha\beta\gamma t}=\partial_\gamma F_{\alpha\beta}
\end{equation}
This is the linearisation of the four-dimensional Lorentzian Taub-NUT solution \cite{NUT, Taub, Misner_NUT} and was referred to as a `gravitypole' in Ref.~\cite{Bunster:2006rt}. This solution was also discussed in Ref.~\cite{Hinterbichler2023GravitySymmetries}.

The dual ADM mass is the charge $\tilde{Q}[\tilde{k}]$ in eq.~\eqref{eq:4d_dual_ADM_charges} with $\tilde{k}$ the constant timelike Killing vector $\tilde{k}^m = \delta^m_t$ and is proportional to the NUT charge $N$ \cite{HullYetAppear} 
\begin{equation}\label{eq:nutty}
    \tilde{Q}[\tilde{k}]= 4\pi N
\end{equation}
We have seen in section~\ref{sec:four_dimensions} that this charge is related to the $\C$-type CKY tensors, e.g. in eq.~\eqref{eq:4d_Q[C]}. To that end, consider a $\C$-type CKY tensor
\begin{equation}\label{eq:LorTaubNUT_CKY}
    K_{mn} = \tilde{\lambda}_{mnp}x^p
\end{equation}
where $\tilde{\lambda}_{mnp}$ has non-zero components
\begin{equation}\label{eq:4d_lambdaconst}
    \tilde{\lambda}_{\alpha\beta\gamma} = - \epsilon_{\alpha\beta\gamma}
\end{equation}
Then $\tilde{k}_m$ is related to $K_{mn}$ by eq.~\eqref{eq:ktilde=*dK}, viz.,
\begin{equation}
    \tilde{k}_m = \frac{1}{3!} \epsilon_{mnpq} \tilde{\lambda}^{npq} \implies \tilde{k}^t = 1  \qc \tilde{k}^\alpha=0
\end{equation}
We evaluate the Penrose charge by integrating over a 2-sphere at constant $r$ and $t$. From eqs.~\eqref{eq:LorTaubNUT_F} and \eqref{eq:LorTaubNUT_Riemann}, we find
\begin{equation}
    Y[K]_{t \alpha} = 2 x_\beta \partial_\alpha \partial_\beta V = - 4 N \frac{x_\alpha}{\abs{x}^3}
\end{equation}
Then integrating over the 2-sphere gives
\begin{equation}\label{eq:Q[K]_LorTaubNUT}
    Q[K] = \int_{S^2} \star Y[K] = 8\pi N 
\end{equation}
The result is independent of $r$ and $t$, reflecting the topological nature of the charge.

Note that the result in eq.~\eqref{eq:Q[K]_LorTaubNUT} is a factor of 2 larger than $\tilde{Q}[\tilde{k}]$ in eq.~\eqref{eq:nutty}. The remaining contribution comes from the topological $\int\dd\star W[\C]$ term in eq.~\eqref{eq:4d_Q[C]}. In particular, for $K$ given by eq.~\eqref{eq:LorTaubNUT_CKY}, we find
\begin{equation}
    \partial^\beta W[\C]_{t\alpha \beta} = -2N \frac{x_\alpha}{\abs{x}^3}
\end{equation}
such that
\begin{equation}\label{eq:topological_TaubNUT}
    \int_{S^2} \dd\star W[\C] = 4\pi N
\end{equation}
Hence the results in eqs.~\eqref{eq:nutty}, \eqref{eq:Q[K]_LorTaubNUT} and \eqref{eq:topological_TaubNUT} are consistent with eq.~\eqref{eq:4d_Q[C]}.

In the linearised gravity theory, this can be simply extended to multi-centred solutions with 
\begin{equation}\label{eq:LorTaubNUT_Xmulti}
    V(x^\alpha) = -\sum_s \frac{N_s}{\abs{x-x_s}}
\end{equation}
for some sources labelled by $s$ of strength $N_s$ at positions $x_s\in\mathbb{R}^3$.
Provided the 2-sphere on which the charge is defined is large enough so as to contain all the sources, we recover the result \eqref{eq:Q[K]_LorTaubNUT} above with $N$ replaced by $\sum_s N_s$. This can be shown directly via a slightly more involved integration. 

We can also consider this as a part of a higher dimensional Kaluza-Klein type solution. Namely, we consider a solution in $d=4+n$ dimensions on the space $(\mathbb{R}^{1,3}\times T^n)^{\bullet}$ which is defined to be $\mathbb{R}^{1,3}\times T^n$ with the cylinders where $x = x_s \in \mathbb{R}^3$ removed. The coordinates on the full space are denoted $x^\mu = (x^m,y^i)$, as in section~\ref{sec:KaluzaKlein}, with $x^m = (t,x^\alpha)$ as above and the coordinates on $T^n$ are periodic with $y^i\sim y^i+2\pi R_i$.
The solution is given by simply taking the only non-zero components of $h_{\mu\nu}$ to be $h_{\alpha t} = 2A_\alpha(x^\beta)$ where $A_\alpha$ satisfies eqs.~\eqref{eq:F=dA}, \eqref{eq:LorTaubNUT_F}, and \eqref{eq:LorTaubNUT_X} so the higher-dimensional solution is the product of the linearised Lorentzian Taub-NUT space with a torus.

In section~\ref{sec:KK_uplift}, we have seen that for Kaluza-Klein solutions of this type, the ADM and dual ADM charges on $(\mathbb{R}^{1,3})^{\bullet}$ are related to charges $Q[\mathscr{K}]$ via eq.~\eqref{eq:Q[Kscr]=Q[k]+Q[xi]}. Here, $\mathscr{K}_{\mu\nu}$ is a 2-form on $(\mathbb{R}^{1,3}\times T^n)^{\bullet}$ related to $K_{mn}$ by eq.~\eqref{eq:Kscr_K_xi}. This can be verified explicitly by calculating $Q[\mathscr{K}]$ for the Lorentzian Taub-NUT solution above.
The integration surface is a codimension-2 cycle in $d$ dimensions, which we take to be $\Sigma = S^2 \times T^n$ where $S^2$ is a 2-sphere of constant $t$ and $r$ within $(\mathbb{R}^{1,3})^\bullet$. From the higher-dimensional perspective, the charge could then be interpreted as that of an $n$-brane fully wrapping the torus. Similar manipulations to those above yield
\begin{equation}\label{eq:LorTaubNUT_PenroseCharge}
    Q[\mathscr{K}] = \int_{\Sigma_3} \star Y[\mathscr{K}] = 8 \pi N \;\text{vol}_{T^n}
\end{equation}
which is indeed related to $Q[K]$ in eq.~\eqref{eq:Q[K]_LorTaubNUT} by a factor of the volume of the $n$-torus, as expected from eq.~\eqref{eq:Q[Kscr]=Q[k]+Q[xi]}.

\subsection{Linearised Kaluza-Klein monopole}
\label{sec:KKMono}

In Ref.~\cite{HullYetAppear} a solution was considered with a source carrying both mass $m$ and a topological charge $p$. When $m=|p|$ this is a linearisation of the Kaluza-Klein monopole solution. In the linearised theory, this can be regarded as a superposition of a solution with mass $m$ and a solution with topological charge $p$, and further more there are superpositions of such solutions with multiple sources at different locations. In this subsection, we will consider multi-source solutions with magnetic charges and show that these carry the charges defined on 2-cycles introduced in section~\ref{sec:d-3_form_charges_S1}.

Consider a background spacetime given by $\mathbb{R}^{1,3}\times S^1$ with the cylinder $\{0\in \mathbb{R}^3\}$ excluded. We denote this space by $(\mathbb{R}^{1,3} \times S^1)^\bullet$ and its coordinates by $x^\mu = (x^m,y)$. The coordinates on $\mathbb{R}^{1,3}$ are $x^m = (t,x^\alpha)$ as in the previous subsection, and the coordinate on the $S^1$ is $y\sim y+2\pi R_y$. We take a Kaluza-Klein Ansatz
\begin{equation}
    h_{\alpha y} = 2A_\alpha(x^\beta)
\end{equation}
where $A_\alpha$ is a 1-form connection on $\mathbb{R}^3\setminus\{0\}$ whose field strength \eqref{eq:F=dA} satisfies
\begin{equation}
    F_{\alpha\beta} = \epsilon_{\alpha\beta\gamma}\partial_\gamma U
\end{equation}
with
\begin{equation}
    U(x^\alpha) = - \frac{p}{\abs{x}}
\end{equation}
This is a similar solution to that of section~\ref{sec:LorTaubNut}, with the roles of $t$ and $y$ reversed. 
Upon reduction over the $S^1$ fibre, this gives a four-dimensional solution with a Dirac monopole of strength $p$ at $x=0$. The five-dimensional Riemann tensor has non-zero components
\begin{equation}
    R_{\alpha\beta\gamma y} = \partial_\gamma F_{\alpha\beta} 
\end{equation}
In particular, for a single monopole at the origin, we can take the Dirac monopole solution of $F_{\theta\phi} = p\sin\theta$ and all other components equal to zero.
Note that $R_{[\alpha\beta\gamma] y} = \partial_{[\gamma} F_{\alpha\beta]}$ is a delta-function at the point $x=0$, so we exclude this point from $\mathbb{R}^{3}$ and exclude the corresponding cylinder $x=0$ from $\mathbb{R}^{1,3}\times S^1$ to give $(\mathbb{R}^{1,3}\times S^1)^\bullet$.

From the discussion in section~\ref{sec:d-3_form_charges_S1}, for Kaluza-Klein type solutions, we can construct the charges $q[\bar{\sigma}]$ and $q[\xi]$ where $\bar{\sigma}_{mn}$ is a closed CKY tensor and $\xi^{(i)}$ are closed CKVs in $D=4$ dimensions.
For the linearised solution above $h_{mn}=0$, so $q[\bar{\sigma}]=0$. However, we find
\begin{equation}\label{eq:d-3_charge_KKmono}
    q[\xi] = \int_{\Sigma_2} X[\xi] = -\frac{1}{2} \hat{\xi} \int_{\Sigma_2} F = 2\pi m p
\end{equation}
where we have written the closed CKV as $\xi_m = n_m + m x_m$ as in eq.~\eqref{eq:xi_solution}, whose divergence is $\hat{\xi}=m$ from eq.~\eqref{eq:xihat}. Here we have taken $\Sigma_2$ to be a 2-sphere at fixed $r=(x_\alpha x^\alpha)^{1/2}$, $t$, and $y$. The fact that the charge $q[\xi]$ is topological (i.e. unchanged by small deformations of the surface on which it is defined) is manifest as the result is independent of $r$, $t$, and $y$.

Again this can be generalised simply in the linearised theory to a solution with multiple sources of strengths $p_s$ at locations $x_s\in\mathbb{R}^3$. In this case all the locations $x_s$ of the sources should be removed from the background manifold and the potential is
\begin{equation}
    U(x^\alpha) = -\sum_s \frac{p_s}{\abs{x-x_s}}
\end{equation}
The result in eq.~\eqref{eq:d-3_charge_KKmono} is modified simply by replacing $p$ by $\sum_s p_s$, provided that $\Sigma_2$ is large enough such that $r>\abs{x_s}$ for all $s$.
The manipulations leading to this result are similar to those that lead to eq.~\eqref{eq:LorTaubNUT_PenroseCharge}.

From eq.~\eqref{eq:d-3_charge_KKmono}, for this example the charge reduces to the first Chern number of the graviphoton field $A_m = h_{my}/2$ evaluated on $\Sigma_2$. Therefore, only when $h$ (and therefore $A$) is a non-globally defined gauge field configuration will these charges be non-zero. The fact that the charge is non-zero and proportional to $p$ in the example above reflects the non-trivial winding of the gauge field configuration around the compact direction. 

\subsection{Linearised C-metric solution}

As noted in section~\ref{sec:BreakdownOfPenroseCharges_4d}, since the constant $\A$-type CKY tensors neither contribute to Killing vectors $\hat{K}$ nor to closed CKY tensors $\tilde{K}$, the Penrose charges $Q[\A]$ are not related to the ADM charges.

An example of a solution with non-zero Penrose charge $Q[\A]$ has been given in Ref.~\cite[eq.~(4.89)]{Hinterbichler2023GravitySymmetries}, where it was said to arise as a linearisation of the C-metric solution to general relativity. 
It has a discontinuity along the $z$-axis and this discontinuity can be remedied by the addition of a pure gauge solution for $z>0$ and a different one for $z<0$, with the two solutions related by a gauge transformation.

This example and the example studied in section~\ref{sec:LorTaubNut} are in accordance with our conclusion that \emph{the Penrose charges correspond to other charges as well as the ADM charges, and that the total derivative term in eq.~\eqref{eq:ADM_Electric_Charges} must be included for topologically non-trivial gauge field configurations}.

\section{Conclusion and outlook}
\label{sec:conclusion}

We have seen that for any CKY tensor $K$, the corresponding improved Penrose 2-form current $Y_+[K]$ is conserved in regions without sources so that $\star Y_+[K]$ is a closed $(d-2)$-form that can be integrated over a $(d-2)$-cycle $\Sigma$ contained in a region where $R_{\mu\nu}=0$ to give a charge $Q[K]$. This gives a topological charge that depends only on the cohomology class of $\star Y_+[K]$ and the homology of $\Sigma$. To obtain non-trivial charges, the space on which the graviton field is  defined cannot be the whole of Minkowski space but must be Minkowski space with some regions removed, so that $\Sigma$ can have non-trivial homology. The excluded regions are associated with the locations of sources.

In Minkowski space, the CKY tensors are given in terms of constant forms $\A$, $\B$, $\C$, $\D$ by eq.~\eqref{eq:CKY_Solution}. 
If $d>4$ then $\star Y_+[\A]$ and $\star Y_+[\C]$ define trivial cohomology classes and so the charges $Q[\A]$ and $Q[\C]$ vanish. They also vanish in $d=4$ if the graviton field is globally defined.
The remaining Penrose charges $Q[\B]$ and $Q[\D]$ then give the standard ADM momentum and angular momentum when the graviton field is globally defined. In the case where it is not globally defined, these charges give improved gauge-invariant versions of the ADM charges.

The most interesting case is that in which $d=4$ and the graviton is not globally defined. Then $Q[\B]$ and $Q[\D]$ give the covariantised ADM momentum and angular momentum as before, but now $Q[\A]$ and $Q[\C]$ can be non-zero and give magnetic-type charges for the graviton. Each magnetic charge can be expressed as the integral of a closed form which is locally exact; however it is the exterior derivative of a form which is not gauge-invariant and so is not globally defined in general. The charge $Q[\C]$ gives the NUT momentum of the linearised theory while the charge $Q[\A]$ is the electric charge for the gauge potential $a_\mu$ defined in eq.~\eqref{eq:a_1form}. In the case in which the 2-form $\A$ is basic, i.e. it defines a 2-plane, then $a_\mu$ can be thought of as the projection of the connection $\Gamma$ onto that 2-plane. 

As a particularly interesting application of our findings, we have considered the Kaluza-Klein setting, in which $d$-dimensional Minkowski space is replaced with the product of $D$-dimensional Minkowski space with a torus of dimension $d-D$. With an appropriate Kaluza-Klein Ansatz to reduce from $d$ to $D$ dimensions, we have found Penrose charges together with graviphoton electric charges in $D$ dimensions.
In this case, interesting gravitational magnetic charges arise for $D=4$.

The Penrose charges may be regarded as generators of $1$-form symmetries while the dual charges given by integrating the Hodge duals of the Penrose currents may be regarded as generators of $(d-3)$-form symmetries.
We have checked that the number of 1-form symmetries equals the number of $(d-3)$-form symmetries in Minkowski space,  in accordance with the discussion of Refs.~\cite{CasiniCompleteness, Benedetti2023GeneralizedGravitons}, and show that this remains true on the product of Minkowski space with a torus.

We have presented a unified framework for discussing charges in linearised gravity and the corresponding currents. In particular, global properties and dualities are discussed. Although some of the results have appeared previously, they are understood here in a wider context that facilitates generalisations.
In particular, the triviality of Penrose charges associated with Killing-Yano tensors in dimensions $d>4$ was shown in Ref.~\cite{Benedetti2023GeneralizedGravitons}. We have extended the relation of Ref.~\cite{Benedetti2023GeneralizedGravitons} to an off-shell identity and gave a systematic construction using the properties of conformal Killing-Yano tensors. 
The equality between the numbers of 1-form and $(d-3)$-form symmetries is expected  on general grounds in Ref.~\cite{CasiniCompleteness} and was seen explicitly for linearised gravity on Minkowski space in Ref.~\cite{Benedetti2023GeneralizedGravitons}. Here we confirm their results and extend them to Kaluza-Klein compactifications of the linearised graviton theory. 
A relation between the Penrose charges and the ADM charges was anticipated in Ref.~\cite{Penrose1982Quasi-localRelativity} and seen in the examples studied in Ref.~\cite{Hinterbichler2023GravitySymmetries}, but the general relation given here is novel and makes a number of properties explicit.

In addition to the charges discussed here, the linear graviton theory also has the magnetic charges of Ref.~\cite{HullYetAppear} given by integrating the currents \eqref{eq:div_of_3_form} with $Z$ given by eq.~\eqref{eq:Z[kappa]}, constructed from $\kappa$-tensors satisfying eq.~\eqref{kkill}. These do not arise from the Penrose charges and it would be interesting to find a covariant origin for them similar to the Penrose construction.

A natural issue is the generalisation of our discussion here to the non-linear theory to give covariant charges for general relativity and supergravity. This will be addressed in a forthcoming paper.

\paragraph{Acknowledgements.}
CH was supported by the STFC Consolidated Grant ST/T000791/1.
MLH was supported by a President's Scholarship from Imperial College London.
UL gratefully acknowledges a Leverhulme visiting professorship to Imperial College as well as the hospitality of the theory group at Imperial.

\appendix

\section{Conformal Killing-Yano tensors}
\label{app:AppendixCKY}

\subsection{General properties}

Consider a manifold $\mathcal{M}$ with metric $g$. A \emph{conformal Killing-Yano (CKY) tensor}, $K$, of rank $p$ is a $p$-form which satisfies
\begin{equation}\label{eq:CKY_definition}
    \nabla_\mu K_{\nu_1\dots\nu_p} = \nabla_{[\mu} K_{\nu_1\dots\nu_p]} + p g_{\mu[\nu_1}\hat{K}_{\nu_2\dots\nu_p]}
\end{equation}
where $\nabla$ is the Levi-Civita connection for $g$ and
\begin{equation}\label{eq:Shat}
    \hat{K}_{\nu_2\dots\nu_p} := \frac{1}{d-p+1} \nabla^{\nu_1}K_{\nu_1\nu_2\dots\nu_p}
\end{equation}
We note that $\hat{K}$ satisfies
\begin{equation}\label{eq:div_Shat_property}
    \nabla^{\nu_2}\hat{K}_{\nu_2\dots\nu_p}=0
\end{equation}
and a further integrability condition of eq.~\eqref{eq:CKY_definition} is that the CKY $K$ satisfies \cite{Lindstrom2022Geometrycurrents}
\begin{equation}\label{eq:two_cov_derivs_on_CKY}
\begin{split}
    \nabla_\mu \nabla_\lambda K_{\nu_1\dots\nu_p} &= (-1)^{p+1}\frac{(p+1)}{2} \mathcal{R}\indices{^\sigma_{\mu[\lambda\nu_1}}K_{\nu_2\dots\nu_p]\sigma} - (p+1) g_{\mu[\lambda}\nabla_{\nu_1}\hat{K}_{\nu_2\dots\nu_p]} \\
    &\qquad+ p \nabla_\mu (g_{\lambda[\nu_1} \hat{K}_{\nu_2\dots\nu_p]})
\end{split}
\end{equation}
where $\mathcal{R}_{\mu\nu\rho\sigma}$ is the curvature tensor for $\nabla$.\footnote{This is not to be confused with $R_{\mu\nu\alpha\beta}$, which we reserve for the linearised Riemann tensor around Minkowski space.}
For the case of $p=2$, it follows from eq.~\eqref{eq:two_cov_derivs_on_CKY} that \cite{Lindstrom2022Killing-YanoCurrents} 
\begin{equation}\label{eq:KV_integrability}
    {(d-2)}\nabla_{(\mu}\hat K_{\nu)}=\mathcal{R}\indices{^\sigma_{(\mu}} K_{\nu)\sigma}
\end{equation}
where $\mathcal{R}_{\mu\nu}$ is the Ricci tensor of the full spacetime. Similar integrability conditions follow for higher-rank CKY tensors.

Two further definitions of note are:
\begin{itemize}
    \item A \emph{Killing-Yano (KY) tensor} is a CKY which is co-closed; that is, $\hat{K}_{\nu_2\dots\nu_p}=0$, hence the final term in eq.~\eqref{eq:CKY_definition} vanishes.
    \item A \emph{closed conformal Killing-Yano (closed CKY) tensor} is a CKY which is closed; hence the first term in eq.~\eqref{eq:CKY_definition} vanishes. 
\end{itemize}
Throughout the discussion, the following facts will be of frequent importance \cite{Howe2018SCKYT, Frolov2008HigherdimensionalVariables}:
\begin{itemize}
    \item The Hodge dual of a CKY tensor is a CKY tensor.
    \item The Hodge dual of a KY tensor is a closed CKY tensor, and vice-versa.
\end{itemize}

A CKY tensor of rank 1 is simply a conformal Killing vector and, similarly, a KY tensor of rank 1 is a Killing vector.
Not all manifolds admit Killing vectors and even fewer admit CKY tensors, although there are examples of spaces with no isometries which admit a rank-2 KY tensor \cite{Dietz1981Space-TimesI}. 

\subsection{CKY tensors of Minkowski space}
\label{app:CKY_Minkowski}

We now describe the CKY tensors of Minkowski space, $\mathcal{M} = \mathbb{R}^{1,d-1}$, though all these results apply to flat space of any metric signature. 
Consider a CKY 2-tensor $K$, which satisfies eq.~\eqref{eq:CKY_equation}. An integrability condition of the CKY equation is that $K$ must satisfy $\partial_\mu \partial_\nu \partial_\rho K_{\alpha\beta}=0$. This is seen as follows. The Riemann curvature tensor vanishes for Minkowski space and hence the integrability condition in eq.~\eqref{eq:KV_integrability} becomes
\begin{equation}
    \partial_{(\mu} \hat{K}_{\nu)} = 0
\end{equation}
so $\hat{K}_\mu$ is a Killing vector on Minkowski space. That is, the general solution for $\hat{K}_\mu$ is simply
\begin{equation}
    \hat{K}_\mu = u_\mu + v_{\mu\nu}x^\nu
\end{equation}
where $u$ is a constant 1-form and $v$ is a constant 2-form. Noting that $\hat{K}_\mu$ is proportional to the divergence of $K_{\mu\nu}$, this implies that $K_{\mu\nu}$ is at most quadratic in $x^\mu$.
Inserting the most general quadratic Ansatz quickly leads to the conclusion that the most general solution is that given in eq.~\eqref{eq:CKY_Solution}. 

Analogous arguments hold for higher-rank CKY tensors on flat space. Namely, integrability conditions of eq.~\eqref{eq:CKY_definition} on flat space imply that $\hat{K}_{\mu_1\dots\mu_{p-1}}$ is a rank-$(p-1)$ KY tensor. Just as the Killing vectors, these are at most linear in $x^\mu$ and so one again finds that the CKY tensors $K_{\mu_1\dots\mu_p}$ are at most quadratic in $x^\mu$. Substituting the most general quadratic Ansatz into the CKY equation then implies that a CKY tensor of rank $p$ can be parameterised by four constant forms, denoted $\A,\B,\C$ and $\D$, of rank $p$, $p+1$, $p-1$ and $p$ respectively, as 
\begin{align}
\begin{split}
    K_{\mu_1\dots\mu_p} &= \A_{\mu_1\dots\mu_p} + \B_{[\mu_1\dots\mu_{p-1}} x_{\mu_p]} + \C_{\mu_1\dots\mu_{p+1}} x^{\mu_{p+1}} \\
    &\phantom{=}+ \frac{1}{2} \D_{\mu_1\dots\mu_p}x^2 + (-1)^p p \D_{\sigma[\mu_1\dots\mu_{p-1}} x_{\mu_p]}x^\sigma \label{eq:flat_space_CKY}
\end{split}
\end{align}
which reduces to eq.~\eqref{eq:CKY_Solution} when $p=2$.

The constant tensors $\A$, $\B$, $\C$, and $\D$ furnish representations of the Lorentz algebra $so(1,d-1)$ which can be assembled into a $(p+1)$-form representation of the conformal algebra $so(2,d)$ \cite{Howe2018SCKYT}. 
Therefore we can interpret the four types of CKY tensors as follows: the $\A$-type term corresponds to translations, the $\B$-type to dilatations, the $\C$-type to rotations, and the $\D$-type to special conformal transformations. 
The $\A$- and $\B$-type terms of eq.~\eqref{eq:flat_space_CKY} give the parameterisation of the closed CKY tensors of flat space, while the $\A$- and $\C$- terms parameterise the KY tensors.

On maximally symmetric spaces with non-zero scalar curvature, there is a unique decomposition of a CKY tensor into a KY tensor and a closed CKY tensor \cite{Tachibana1969OnSpace}. On flat space, however, this decomposition degenerates because the term involving $\A$ in eq.~\eqref{eq:flat_space_CKY} is both closed and co-closed, also the $\D$-type solution is neither closed nor co-closed.

We now explicitly verify that the dual of a rank-2 CKY tensor on four-dimensional Minkowski space is also a CKY tensor. Defining the dual by
\begin{equation}
	(\star K)_{\mu\nu} = \frac{1}{2!} \epsilon_{\mu\nu\alpha\beta} K^{\alpha\beta}
\end{equation}
and taking the solutions on flat space in eq.~\eqref{eq:CKY_Solution}, one explicitly finds
\begin{equation}
	(\star K)_{\mu\nu} = \tilde{\A}_{\mu\nu} - \frac{1}{2}\tilde{\B}_{\mu\nu\rho} x^\rho + 2 \tilde{\C}_{[\mu}x_{\nu]} - \left( 2 x_{[\mu}\tilde{\D}_{\nu]\rho}x^\rho + \frac{1}{2} \tilde{\D}_{\mu\nu} x^2 \right)
\end{equation}
where the Hodge duals of the constant coefficients $\A$, $\B$, $\C$, and $\D$ are defined as
\begin{equation}
	\tilde{\A}_{\mu\nu} = \frac{1}{2!}\epsilon_{\mu\nu\alpha\beta} \A^{\alpha\beta} \qc \tilde{\B}_{\mu\nu\rho} = \epsilon_{\mu\nu\rho\alpha} \B^\alpha \qc \tilde{\C}_\mu = \frac{1}{3!} \epsilon_{\mu\alpha\beta\gamma} \C^{\alpha\beta\gamma} \qc \tilde{\D}_{\mu\nu} = \frac{1}{2!} \epsilon_{\mu\nu\alpha\beta} \D^{\alpha\beta}
\end{equation}
Therefore, the dual $\star K$ is also a CKY tensor, parameterised by the same four tensors $\A$, $\B$, $\C$ and $\D$ as $K$. We note that the terms involving the $\B$ and $\C$ coefficients transform into each other under duality, whereas the $\A$ and $\D$ terms transform into themselves (up to a sign). That is to say, KY tensors are dual to closed CKY tensors and vice-versa (this also holds on general manifolds). These results hold on Minkowski space of any dimension.

We now show that the divergence of a rank-2 CKY tensor on flat space gives a Killing vector and the curl gives a closed CKY tensor. The divergence of a rank-2 CKY tensor on flat space gives
\begin{equation}
	\partial^\beta K_{\alpha\beta} = \frac{d-1}{2} \left( \B_\alpha - 2\D_{\alpha\beta}x^\beta \right)
\end{equation}
The $\B$-type terms give rise to translational Killing vectors, and the $\D$-type ones give rotational ones.
Similarly, the curl of a rank-2 CKY tensor gives
\begin{equation}\label{2to3}
    \partial_{[\alpha}K_{\beta\gamma]} = \C_{\alpha\beta\gamma} + 3\D_{[\alpha\beta} x_{\gamma]}
\end{equation}
which is precisely the form of a rank-3 closed CKY tensor. These results generalise to CKY tensors of arbitrary rank on Minkowski space. Namely, the divergence of a rank-$p$ CKY tensor is a rank-$(p-1)$ KY tensor, while its curl is a rank-$(p+1)$ closed CKY tensor. Furthermore, on flat space all KY tensors can be written as the divergence of a CKY tensor, and all closed CKY tensors can be written as the curl of a CKY tensor.

\subsection{Relations with dual tensors}
\label{app:dualCKYs}

We have shown that given a rank-2 CKY tensor $K$ on flat space, $\hat{K}_{\mu} = \frac{1}{d-1} \partial^\nu K_{\nu\mu}$ is a Killing vector and $\tilde{K}_{\mu\nu\rho} = \partial_{[\mu}K_{\nu\rho]}$ is a closed CKY 3-form. It follows that $\star\hat{K}$ is a closed CKY tensor of rank $d-1$ and $\star\tilde{K}$ is a KY of rank $n\equiv d-3$. We now derive the relations between these objects.

We repeat the CKY condition in eq.~\eqref{eq:CKY_equation} here for reference,
\begin{equation} \label{eq:CKY_equation_k_lambda_repeat}
    \partial_\alpha K_{\beta\gamma} = \tilde{K}_{\alpha\beta\gamma} +2 \eta_{\alpha[\beta} \hat{K}_{\gamma]}
\end{equation}
Contracting this with $\partial_\beta$, we have
\begin{align}
    \partial^\alpha \partial_\beta K^{\beta\gamma} = - \partial_\beta \tilde{K}^{\beta\alpha\gamma} +2 \eta^{\alpha[\beta}\partial_\beta \hat{K}^{\gamma]}
\end{align}
which implies, using $\partial_\beta \hat{K}^\beta=0$,
\begin{equation}
    \partial_\beta \tilde{K}^{\beta\alpha\gamma} = -(d-2) \partial^\alpha \hat{K}^\gamma \label{eq:Fact.divlambdatilde=dk}
\end{equation}

Similar considerations can be done for the duals. Firstly, contracting eq.~\eqref{eq:CKY_equation_k_lambda_repeat} with $\epsilon^{\mu_1\dots\mu_{n+1}\beta\gamma}$ gives
\begin{align}
    \frac{1}{2} \epsilon^{\mu_1\dots\mu_{n+1}\beta\gamma} \partial^\alpha K_{\beta\gamma} &= \frac{1}{2} \epsilon^{\mu_1\dots\mu_{n+1}\beta\gamma} \left(\tilde{K}\indices{^\alpha_{\beta\gamma}} +2 \delta^\alpha_{[\beta} \hat{K}_{\gamma]}\right) \nonumber \\
    &= \frac{1}{2!n!} \epsilon^{\mu_1\dots\mu_{n+1}\beta\gamma} \epsilon\indices{^\alpha_{\beta\gamma \sigma_1\dots\sigma_n}} (\star\tilde{K})^{\sigma_1\dots\sigma_n} + \epsilon^{\mu_1\dots\mu_{n+1}\alpha\gamma} \hat{K}_\gamma
\end{align}
which implies
\begin{align}
    \partial^\alpha (\star K)^{\mu_1\dots\mu_{n+1}} = -(n+1) \eta^{\alpha[\mu_1} (\star\tilde{K})^{\mu_2\dots\mu_{n+1}]} + (-1)^{n+1} (\star\hat{K})^{\alpha\mu_1\dots\mu_{n+1}} \label{eq:Working.Ktilde}
\end{align}
where
\begin{align}
    (\star K)^{\mu_1\dots\mu_{n+1}} &= \frac{1}{2!} \epsilon^{\mu_1\dots\mu_{n+1}\beta\gamma} K_{\beta\gamma} \\
    (\star\hat{K})_{\alpha_1\dots\alpha_{n+2}} &= \epsilon_{\alpha_1\dots\alpha_{n+2} \beta} \hat{K}^\beta
\end{align}
Contracting the $\alpha$ and $\mu_1$ indices in eq.~\eqref{eq:Working.Ktilde} gives
\begin{equation}
    \partial_\alpha (\star K)^{\alpha\mu_2\dots\mu_{n+1}} = - 3 (\star\tilde{K})^{\mu_2\dots\mu_{n+1}} \label{eq:Working.divKtilde}
\end{equation}
Now acting on eq.~\eqref{eq:Working.Ktilde} with $\partial_{\mu_1}$ and using eq.~\eqref{eq:Working.divKtilde} and the fact that $\partial_{\mu_1}(\star\tilde{K})^{\mu_1\dots\mu_n}=0$ (since $\star\tilde{K}$ is a KY), one finds 
\begin{equation}
    \partial_\sigma (\star\hat{K})^{\sigma\alpha\mu_2\dots\mu_{n+1}} = 2(-1)^{n+1} \partial^\alpha (\star\tilde{K})^{\mu_2\dots\mu_{n+1}} \label{eq:Fact.dlambda=divktilde}
\end{equation}

\section{Manipulations involving the Penrose 2-form}

\subsection{Relating the Penrose current to the secondary  current $J[k]$}
\label{app:AppendixRiemann}

In this appendix we show the manipulations leading to eq.~\eqref{eq:ADM_ImprovedPenrose_Relation}. Firstly, we manipulate the improved Penrose 2-form to write it in terms of the connection $\Gamma$ of eq.~\eqref{eq:DefConnection}. 

We recall from the discussion in section~\ref{sec:Penrose2Form} that $\tilde{K}_{\mu\nu\rho} = \partial_{[\mu}K_{\nu\rho]}$ is a closed CKY 3-form, and that $\hat{K}_\mu = \frac{1}{d-1}\partial^\nu K_{\nu\mu}$ is a Killing vector.
In terms of $\hat{K}$ and $\tilde{K}$, the CKY equation for $K$ is given in eq.~\eqref{eq:CKY_equation}, which we repeat here for convenience,
\begin{equation} \label{eq:CKY_equation_k_lambda}
    \partial_\mu K_{\nu\rho} = \tilde{K}_{\mu\nu\rho} + 2 \eta_{\mu[\nu} \hat{K}_{\rho]}
\end{equation}
Now, from eq.~\eqref{eq:DefRiemann}, we have
\begin{align}
    K^{\mu\nu} R &= 2K^{\mu\nu} \partial_\alpha \Gamma\indices{^{\alpha\beta}_{|\beta}} \nonumber \\
    &= 2\partial_\alpha \left( K^{\mu\nu} \Gamma\indices{^{\alpha\beta}_{|\beta}} \right) - 2\partial^\alpha K^{\mu\nu} \Gamma\indices{_{\alpha\beta}^{|\beta}} \nonumber \\
    &= 2\partial_\alpha \left( 3 K^{[\mu\nu} \Gamma\indices{^{\alpha]\beta}_{|\beta}} - 2 K^{\alpha[\mu}\Gamma\indices{^{\nu]\beta}_{|\beta}} \right) - 2 \partial^\alpha K^{\mu\nu} \Gamma\indices{_{\alpha\beta}^{|\beta}} \nonumber\\
    &= \partial_\alpha \Lambda_1^{\mu\nu\alpha} -4 \partial_\alpha K^{\alpha[\mu}\Gamma\indices{^{\nu]\beta}_{|\beta}} -4 K^{\alpha[\mu} \partial_\alpha \Gamma\indices{^{\nu]\beta}_{|\beta}} - 2\partial^\alpha K^{\mu\nu} \Gamma\indices{_{\alpha\beta}^{|\beta}} \nonumber\\
    &= \partial_\alpha \Lambda_1^{\mu\nu\alpha} -4 K^{\alpha[\mu} \partial_\alpha \Gamma\indices{^{\nu]\beta}_{|\beta}} -4(d-1) \hat{K}^{[\mu} \Gamma\indices{^{\nu]\beta}_{|\beta}} - 2\left( \tilde{K}^{\alpha\mu\nu} +2 \eta^{\alpha[\mu}\hat{K}^{\nu]} \right) \Gamma\indices{_{\alpha\beta}^{|\beta}} \nonumber\\
    &= \partial_\alpha \Lambda_1^{\mu\nu\alpha} -4 K^{\alpha[\mu} \partial_\alpha \Gamma\indices{^{\nu]\beta}_{|\beta}} - 2\tilde{K}^{\mu\nu\alpha} \Gamma\indices{_{\alpha\beta}^{|\beta}} - 4(d-2) \hat{K}^{[\mu} \Gamma\indices{^{\nu]\beta}_{|\beta}} \label{eq:K.RicciScalar}
\end{align}
where we have integrated by parts in the second equality and then used the CKY equation \eqref{eq:CKY_equation_k_lambda} in order to write the result in terms of $\hat{K}$ and $\tilde{K}$.
We have collected total divergence terms into a 3-form $\Lambda_1$ given by
\begin{equation}
    \Lambda_1^{\mu\nu\alpha} = 6 K^{[\mu\nu} \Gamma\indices{^{\alpha]\beta}_{|\beta}}
\end{equation}
Similarly, we have
\begin{align}
    R^{\mu\nu\alpha\beta}K_{\alpha\beta} &= 2K^{\alpha\beta}\partial_\alpha \Gamma\indices{^{\mu\nu}_{|\beta}} \nonumber \\
    &= 2\partial_\alpha (K^{\alpha\beta}\Gamma\indices{^{\mu\nu}_{|\beta}}) - 2\partial_\alpha K^{\alpha\beta} \Gamma\indices{^{\mu\nu}_{|\beta}} \nonumber \\
    &= -2\partial_\alpha \left( 3 K^{\beta[\alpha}\Gamma\indices{^{\mu\nu]}_{|\beta}} - 2 K^{\beta[\mu} \Gamma\indices{^{\nu]\alpha}_{|\beta}} \right) - 2(d-1) \hat{K}^\beta \Gamma\indices{^{\mu\nu}_{|\beta}} \nonumber \\
    &= \partial_\alpha \Lambda_2^{\mu\nu\alpha} + 4 \partial_\alpha K^{\beta[\mu} \Gamma\indices{^{\nu]\alpha}_{|\beta}} + 4 K^{\beta[\mu} \partial_\alpha \Gamma\indices{^{\nu]\alpha}_{|\beta}} - 2(d-1)\hat{K}^\beta \Gamma\indices{^{\mu\nu}_{|\beta}} \nonumber \\
    &= \partial_\alpha \Lambda_2^{\mu\nu\alpha} + 4 K^{\beta[\mu} \partial_\alpha \Gamma\indices{^{\nu]\alpha}_{|\beta}} - 2(d-1) \hat{K}^\beta \Gamma\indices{^{\mu\nu}_{|\beta}}  + 4\left( \tilde{K}^{\alpha\beta[\mu} + \eta^{\alpha\beta}\hat{K}^{[\mu} - \hat{K}^\beta \eta^{\alpha[\mu} \right) \Gamma\indices{^{\nu]}_{\alpha|\beta}} \nonumber \\
    &= \partial_\alpha \Lambda_2^{\mu\nu\alpha} + 4 K^{\alpha[\mu} \partial_\beta \Gamma\indices{^{\nu]\beta}_{|\alpha}} + 4\tilde{K}^{\alpha\beta[\mu}\Gamma\indices{^{\nu]}_{\alpha|\beta}} - 2(d-3) \hat{K}^\beta \Gamma\indices{^{\mu\nu}_{|\beta}} +4 \hat{K}^{[\mu}\Gamma\indices{^{\nu]\beta}_{|\beta}} \label{eq:K.Riemann}
\end{align}
where we have integrated by parts in the second equality and again used the CKY condition \eqref{eq:CKY_equation_k_lambda}. Again, we have collected total divergence terms into a 3-form $\Lambda_2$ given by
\begin{equation}
    \Lambda_2^{\mu\nu\alpha} = -6 K^{\beta[\mu} \Gamma\indices{^{\nu\alpha]}_{|\beta}}
\end{equation}
The sum of eqs.~\eqref{eq:K.RicciScalar} and \eqref{eq:K.Riemann} is
\begin{equation}
\begin{split}
    R^{\mu\nu\alpha\beta}K_{\alpha\beta} + R K^{\mu\nu} &= \partial_\alpha \Lambda^{\mu\nu\alpha} -4 K^{\alpha[\mu} \left( \partial_\alpha \Gamma\indices{^{\nu]\beta}_{|\beta}} - \partial_\beta \Gamma\indices{^{\nu]\beta}_{|\alpha}} \right) + 4\tilde{K}^{\alpha\beta[\mu}\Gamma\indices{^{\nu]}_{\alpha|\beta}} \\
    & \quad - 2\tilde{K}^{\mu\nu\alpha} \Gamma\indices{_{\alpha\beta}^{|\beta}} - 2(d-3)\left( \hat{K}^\beta \Gamma\indices{^{\mu\nu}_{|\beta}} +2 \hat{K}^{[\mu} \Gamma\indices{^{\nu]\beta}_{|\beta}} \right) \label{eq:SumRiemannRicci}
\end{split}
\end{equation}
where 
\begin{equation}
    \Lambda^{\mu\nu\alpha} = \Lambda_1^{\mu\nu\alpha} + \Lambda_2^{\mu\nu\alpha} = 12 K^{[\mu\nu} \Gamma\indices{^{\alpha\beta]}_{|\beta}}
\end{equation}
It follows from eq.~\eqref{eq:DefRiemann} that the linearised Ricci tensor can be written
\begin{equation}
    R\indices{^\mu_\alpha} = \partial_\alpha \Gamma\indices{^{\mu\beta}_{|\beta}} - \partial_\beta \Gamma\indices{^{\mu\beta}_{|\alpha}}
\end{equation}
Hence eq.~\eqref{eq:SumRiemannRicci} can be written
\begin{align}
\begin{split}
     R^{\mu\nu\alpha\beta}K_{\alpha\beta} + 4R^{\alpha[\mu} K\indices{^{\nu]}_\alpha} + R K^{\mu\nu} &= \partial_\alpha \Lambda^{\mu\nu\alpha} + 4\tilde{K}^{\alpha\beta[\mu}\Gamma\indices{^{\nu]}_{\alpha|\beta}} - 2\tilde{K}^{\mu\nu\alpha} \Gamma\indices{_{\alpha\beta}^{|\beta}} \\
     & \qquad -2(d-3) \left( \hat{K}^\beta \Gamma\indices{^{\mu\nu}_{|\beta}} + 2 \hat{K}^{[\mu} \Gamma\indices{^{\nu]\beta}_{|\beta}} \right) 
\end{split}
\end{align}
The left-hand side is the improved Penrose current in eq.~\eqref{eq:Def_ImprovedPenrose}, and a simple rearrangement of the right-hand side yields
\begin{equation}\label{eq:ImprovedPenroseGammaForm}
    Y_+[K]^{\mu\nu} = \partial_\alpha \Lambda^{\mu\nu\alpha} + 6\tilde{K}^{\alpha[\mu\nu}\Gamma\indices{^{\beta]}_{\alpha|\beta}} -6(d-3) \hat{K}^{[\beta} \Gamma\indices{^{\mu\nu]}_{|\beta}}
\end{equation}

We now show that the final two terms of the right-hand side give the secondary current $J[k]$, with $k=\hat{K}$, up to a total divergence. 
First, consider the term involving $\hat{K}$, we have
\begin{align} \label{eq:k_term_working}
    \hat{K}^{[\beta} \Gamma\indices{^{\mu\nu]}_{|\beta}} = \eta^{\mu\nu\alpha|\rho\sigma\beta} \Gamma_{\sigma\beta|\alpha} \hat{K}_\rho = \eta^{\mu\nu\alpha|\rho\sigma\beta} \partial_\sigma h_{\alpha\beta} \hat{K}_\rho = - \frac{1}{3} \partial_\sigma \mathcal{K}^{\mu\nu|\rho\sigma} \hat{K}_\rho
\end{align}
where $\mathcal{K}^{\mu\nu|\rho\sigma}$ is given in eq.~\eqref{eq:DefK}.
For the term involving $\tilde{K}$ in eq.~\eqref{eq:ImprovedPenroseGammaForm}, we use $\Gamma_{[\alpha\beta|\gamma]}=0$ to write 
\begin{equation}
    6 \tilde{K}^{\alpha[\mu\nu}\Gamma\indices{^{\beta]}_{\alpha|\beta}} = -4 \tilde{K}^{[\mu\nu\alpha} \Gamma\indices{_{\alpha\beta|}^{\beta]}} = -4 \delta^{\mu\nu\alpha\beta}_{\rho\sigma\kappa\tau} \tilde{K}^{\rho\sigma\kappa} \Gamma\indices{_{\alpha\beta|}^{\tau}}
\end{equation}
Now using the identity
\begin{equation}
    \epsilon^{\mu\nu\alpha\beta\gamma_2\dots\gamma_{n}} \epsilon_{\rho\sigma\kappa\tau\gamma_2\dots\gamma_{n}} = - 4! (n-1)! \delta^{\mu\nu\alpha\beta}_{\rho\sigma\kappa\tau}
\end{equation}
where $n\equiv d-3$, we have
\begin{align}
    6 \tilde{K}^{\alpha[\mu\nu}\Gamma\indices{^{\beta]}_{\alpha|\beta}} &= - \frac{1}{(n-1)!} \epsilon^{\mu\nu\alpha\beta\gamma_2\dots\gamma_{n}} \partial_\alpha h\indices{_\beta^\tau} (\star\tilde{K})_{\tau\gamma_2\dots\gamma_{n}} \nonumber \\
    &= \partial_\alpha \Xi^{\mu\nu\alpha} + \frac{1}{(n-1)!} \epsilon^{\mu\nu\alpha\beta\gamma_2\dots\gamma_{n}} h\indices{_\beta^\tau} \partial_\alpha (\star\tilde{K})_{\tau\gamma_2\dots\gamma_{n}}
\end{align}
with
\begin{equation}\label{eq:*tildeK}
    \tilde{K}_{\alpha\beta\gamma} = \frac{1}{n!} \epsilon_{\alpha\beta\gamma\sigma_1\dots\sigma_n} (\star\tilde{K})^{\sigma_1\dots\sigma_n}
\end{equation}
and $\Xi$ is a 3-form given by
\begin{equation}
    \Xi^{\mu\nu\alpha} = - \frac{1}{(n-1)!} \epsilon^{\mu\nu\alpha\beta\gamma_2\dots\gamma_{n}} h\indices{_\beta^\tau} (\star\tilde{K})_{\tau\gamma_2\dots\gamma_{n}} = 4h\indices{_\beta^{[\mu}} \tilde{K}^{\nu\alpha\beta]}
\end{equation}
Note that since $\tilde{K}$ is a closed CKY 3-form, $\star\tilde{K}$ is a KY $(d-3)$-form. Therefore, it satisfies the KY condition $\partial_\alpha (\star\tilde{K})_{\tau\gamma_2\dots\gamma_{n}} = \partial_{[\alpha} (\star\tilde{K})_{\tau\gamma_2\dots\gamma_{n}]}$. Using this, we have
\begin{align} \label{eq:lambda_term_working}
    6 \tilde{K}^{\alpha[\mu\nu}\Gamma\indices{^{\beta]}_{\alpha|\beta}} &= \partial_\alpha \Xi^{\mu\nu\alpha} + \frac{1}{(n-1)!} \epsilon^{\mu\nu\beta\gamma_1\dots\gamma_n} h\indices{_\beta^\tau} \partial_\tau (\star\tilde{K})_{\gamma_1\dots\gamma_n}
\end{align}
It is shown in appendix~\ref{app:dualCKYs} that $\star\tilde{K}$ is non-locally related to $\star\hat{K}$ by 
\begin{equation}
    \partial_\sigma (\star\hat{K})^{\sigma\tau\gamma_1\dots\gamma_n} = 2(-1)^{n+1} \partial^\tau (\star\tilde{K})^{\gamma_1\dots\gamma_n}
\end{equation}
where
\begin{equation}
    (\star\hat{K})_{\alpha_1\dots\alpha_{n+2}} = \epsilon_{\alpha_1\dots\alpha_{n+2}\beta}\hat{K}^\beta
\end{equation}
is a rank-$(d-1)$ closed CKY tensor dual to the Killing vector $\hat{K}$.
Substituting into eq.~\eqref{eq:lambda_term_working} gives
\begin{align}
    6 \tilde{K}^{\alpha[\mu\nu}\Gamma\indices{^{\beta]}_{\alpha|\beta}} &= \partial_\alpha \Xi^{\mu\nu\alpha} - \frac{1}{2(n-1)!} \epsilon^{\mu\nu\beta\gamma_1\dots\gamma_n} \epsilon_{\sigma\tau\rho\gamma_1\dots\gamma_n} h\indices{_\beta^\tau} \partial^\sigma \hat{K}^\rho \nonumber \\
    &= \partial_\alpha \Xi^{\mu\nu\alpha} -3(d-3) \eta^{\mu\nu\beta|\sigma\rho\tau} h_{\beta\tau} \partial_\sigma \hat{K}_\rho \nonumber \\
    &= \partial_\alpha \Xi^{\mu\nu\alpha} +(d-3) \mathcal{K}^{\mu\nu|\sigma\rho} \partial_\sigma \hat{K}_\rho \label{eq:lambda_term_working2}
\end{align}
Now from $\eta^{[\alpha\sigma\mu|\nu]\rho\beta}=0$ we note that
\begin{equation} \label{eq:K_irreducible}
    \mathcal{K}^{[\sigma\mu|\nu]\rho} = -3 \eta^{\alpha[\sigma\mu|\nu]\rho\beta} h_{\alpha\beta} = \eta^{\sigma\mu\nu|\alpha\beta\rho} h_{\alpha\beta} = 0
\end{equation}
Hence
\begin{equation}
    \mathcal{K}^{\mu\nu|\sigma\rho} + 2\mathcal{K}^{\sigma[\mu|\nu]\rho} = 0
\end{equation}
Substituting this into eq.~\eqref{eq:lambda_term_working2} we find
\begin{equation} \label{eq:lambda_term_working3}
    6 \tilde{K}^{\alpha[\mu\nu} \Gamma\indices{^{\beta]}_{\alpha|\beta}} = \partial_\alpha \Xi^{\mu\nu\alpha} -2(d-3) \mathcal{K}^{\sigma\mu|\nu\rho} \partial_\sigma \hat{K}_\rho 
\end{equation}
Note that the result is antisymmetric in the $\mu$ and $\nu$ indices as a result of eq.~\eqref{eq:K_irreducible} and the fact that $\hat{K}$ is a Killing vector.
Finally, substituting the results of eqs.~\eqref{eq:k_term_working} and \eqref{eq:lambda_term_working3} into eq.~\eqref{eq:ImprovedPenroseGammaForm} we find
\begin{align}
    Y_+[K]^{\mu\nu} &= 2(d-3) \left( \partial_\sigma \mathcal{K}^{\mu\nu|\rho\sigma} \hat{K}_\rho - \mathcal{K}^{\sigma\mu|\nu\rho} \partial_\sigma \hat{K}_\rho \right) + \partial_\rho (\Lambda^{\mu\nu\rho} + \Xi^{\mu\nu\rho}) \nonumber \\
    &= 2(d-3) J[\hat{K}]^{\mu\nu} + \partial_\rho Z^{\mu\nu\rho} 
\end{align}
where the result follows from comparison with eq.~\eqref{eq:J[k]_def}, and $Z^{\mu\nu\rho} = \Lambda^{\mu\nu\rho} + \Xi^{\mu\nu\rho}$. Therefore, associating the Killing vector $k$ with the divergence of the CKY tensor $K$,
\begin{equation}
    k_\mu = 2(d-3) \hat{K}_\mu
\end{equation}
we find eq.~\eqref{eq:ADM_ImprovedPenrose_Relation}.

\subsection{Conditions for the closure of the Penrose 2-form}
\label{app:closure_Y}

We have seen in section~\ref{sec:d-3_form_charges} that in $d>4$ the Penrose 2-form $Y[K]$ is closed on-shell in regions without sources only when $K$ is a closed CKY tensor.
In this appendix we show that in $d=4$, $Y[K]$ is closed for all CKY tensors in such regions. Firstly, we recall eq.~\eqref{eq:dY_working},
\begin{equation}
    \partial_{[\rho} Y[K]_{\mu\nu]} = R_{\alpha\beta[\mu\nu} \tilde{K}\indices{_{\rho]}^{\alpha\beta}}
\end{equation}
This can equivalently be written
\begin{align}
    \partial_{[\rho} Y[K]_{\mu\nu]} &= \delta^{\sigma\kappa\lambda}_{\mu\nu\rho} R_{\alpha\beta\sigma\kappa} \tilde{K}\indices{_{\lambda}^{\alpha\beta}} \nonumber \\
    &= -\frac{1}{3!} \epsilon_{\mu\nu\rho\gamma} \epsilon^{\sigma\kappa\lambda\gamma} R_{\alpha\beta\sigma\kappa} \tilde{K}\indices{_\lambda^{\alpha\beta}}
\end{align}
where $\delta^{\sigma\kappa\lambda}_{\mu\nu\rho} = \delta^\sigma_{[\mu} \delta^\kappa_\nu \delta^\lambda_{\rho]}$. In the second equality we have used that
\begin{equation}\label{eq:ep_ep_delta}
    \epsilon_{\mu\nu\rho\gamma} \epsilon^{\sigma\kappa\lambda\gamma} = - 3! \delta^{\sigma\kappa\lambda}_{\mu\nu\rho}
\end{equation}
The Hodge dual satisfies
\begin{equation}
    \tilde{K}_{\lambda\alpha\beta} = \epsilon_{\lambda\alpha\beta\delta} (\star \tilde{K})^\delta
\end{equation}
which, along with eq.~\eqref{eq:ep_ep_delta}, implies
\begin{align}
    \partial_{[\rho} Y[K]_{\mu\nu]} &= -\epsilon_{\mu\nu\rho\gamma} \delta^{\sigma\kappa\gamma}_{\alpha\beta\delta} R\indices{^{\alpha\beta}_{\sigma\kappa}} (\star\tilde{K})^\delta \nonumber \\
    &= \frac{2}{3} \epsilon_{\mu\nu\rho\gamma} G\indices{^\gamma_\delta} (\star\tilde{K})^\delta \label{eq:dY_working_4d_appendix}
\end{align}
where we have used that
\begin{equation}
    G\indices{^\mu_\nu} = -\frac{3}{2} \delta^{\mu\alpha\beta}_{\nu\gamma\delta} R\indices{_{\alpha\beta}^{\gamma\delta}}
\end{equation}
in the final equality. From eq.~\eqref{eq:dY_working_4d_appendix}, we see that $Y[K]$ is closed on-shell for all CKY tensors in regions without sources. If this series of manipulations are used in $d>4$, the result does not give the Einstein tensor and does not vanish on-shell.

\bibliographystyle{JHEP}
\bibliography{references}

\end{document}